\documentclass[oneside,english]{siamltex}
\usepackage[T1]{fontenc}
\usepackage[latin9]{inputenc}
\usepackage{babel}
\usepackage{float}
\usepackage{amstext}
\usepackage{graphicx}
\usepackage{esint}
\usepackage[unicode=true,
 bookmarks=true,bookmarksnumbered=false,bookmarksopen=false,
 breaklinks=false,pdfborder={0 0 0},backref=false,colorlinks=false]
 {hyperref}
\hypersetup{pdftitle={FluSI: A novel parallel simulation tool for flapping insect flight using a Fourier method with volume penalization},
 pdfauthor={T. Engels et al.}}

\makeatletter

\floatstyle{ruled}
\newfloat{algorithm}{tbp}{loa}
\providecommand{\algorithmname}{Algorithm}
\floatname{algorithm}{\protect\algorithmname}

\newcommand\eqref[1]{(\ref{#1})}

\usepackage{algorithm,algpseudocode}

\makeatother

\begin{document}

\title{FluSI: A novel parallel simulation tool for flapping insect flight
using a Fourier method with volume penalization\footnote[5]{
T\lowercase{his work was granted access to the}
HPC 
\lowercase{resources of}
A\lowercase{ix}-
M\lowercase{arseille} 
U\lowercase{niversit\'e financed by the project Equip$@$Meso}
(ANR-10-EQPX-29-01) 
\lowercase{of the program « Investissements d'Avenir » supervised by the}
A\lowercase{gence} N\lowercase{ationale pour la} R\lowercase{echerche, as well as to the}
HPC
\lowercase{resources of }
IDRIS 
(I\lowercase{nstitut du} d\lowercase{éveloppement et des } r\lowercase{essources en }i\lowercase{nformatique} s\lowercase{cientifique) under project number i20152a1664.}
TE,KS,JS 
\lowercase{thank the } f\lowercase{rench}-g\lowercase{erman} u\lowercase{niversity for financial support.} 
T\lowercase{he authors wish to thank} M\lowercase{asateru} M\lowercase{aeda for fruitful discussions.}
}}

\author{Thomas Engels\footnote[1]{C\lowercase{orresponding author, e-mail address: thomas.engels$@$mailbox.tu-berlin.de}} 
\footnote[2]{
ISTA,
T\lowercase{echnische} 
U\lowercase{niversit\"at} 
B\lowercase{erlin},
M\lowercase{\"uller}-B\lowercase{reslau}-S\lowercase{trasse} 12, 
10623 B\lowercase{erlin}, G\lowercase{ermany} 
}
\footnotemark[3], Dmitry Kolomenskiy\footnote[4]{
B\lowercase{iomechanical} 
E\lowercase{ngineering} 
L\lowercase{aboratory},
C\lowercase{hiba} U\lowercase{niversity}, 1-33, 
Y\lowercase{ayoi-cho}, I\lowercase{nage-ku}, C\lowercase{hiba-shi}, C\lowercase{hiba}, 263-8522, J\lowercase{apan}}, Kai Schneider\footnote[3]{
M2P2 {UMR}7340, CNRS 
\lowercase{and}
A\lowercase{ix}-M\lowercase{arseille} 
U\lowercase{niversit\'e},
38 \lowercase{rue} 
J\lowercase{oliot}-C\lowercase{urie}
13451
M\lowercase{arseille} \lowercase{cedex} 20, F\lowercase{rance}} and Jörn Sesterhenn\footnotemark[2]}
\maketitle
\begin{abstract}
FluSI, a fully parallel open source software for pseudo-spectral simulations
of three-dimensional flapping flight in viscous flows, is presented.
It is freely available for non-commercial use under \href{https://github.com/pseudospectators/FLUSI}{https://github.com/pseudospectators/FLUSI}.
The computational framework runs on high performance computers with
distributed memory architectures. The discretization of the three-dimensional
incompressible Navier--Stokes equations is based on a Fourier pseudo-spectral
method with adaptive time stepping. The complex time varying geometry
of insects with rigid flapping wings is handled with the volume penalization
method. The modules characterizing the insect geometry, the flight
mechanics and the wing kinematics are described. Validation tests
for different benchmarks illustrate the efficiency and precision of
the approach. Finally, computations of a model insect in the turbulent
regime demonstrate the versatility of the software. \end{abstract}



\section{Introduction}

Flapping flight is an active interdisciplinary research field with
many open questions, and numerical simulations have become an important
instrument for tackling them. Here, we present the computational solution
environment \texttt{FluSI} which is, to the best of our knowledge,
the first open source code in this field.

In the literature different numerical strategies have been proposed
simulating flapping flight, e.g., the use of several, overlapping
grids \cite{Liu2009}, that are adapted to the geometry, or the use
of moving grids with the Arbitrary Lagrangian-Eulerian method \cite{Ramamurti2009}.
Those methods involve significant computational and implementation
overhead, whose reduction has motivated the development of methods
that allow for a geometry-independent discretization, such as immersed
boundary methods. 

In this work, we employ the volume penalization method to take the
no--slip boundary condition into account. The idea of modeling solid
obstacles as porous media with a small permeability has been proposed
in \cite{Arquis1984}. The forcing term acts on the entire volume
of the solid and not only on its surface, as it is the case in the
immersed boundary methods, and corresponds to the Darcy drag. The
distinctive feature of the method is the existence of rigorous convergence
proofs \cite{Angot1999,Carbou2003} showing that the solution of the
penalized equations does indeed converge to the solution of the Navier-Stokes
equations with no-slip boundary conditions. This approach has been
extended to model not only Dirichlet conditions applied at the surface
of moving, rigid and flexible obstacles \cite{Kolomenskiy2009,Engels2014},
but also to homogeneous Neumann conditions, which is relevant for
studying, e.g., turbulent mixing of a passive scalar \cite{Kadoch_etal_2012}.
As animals forage following odor traces, this technique can potentially
be attractive for insect flight simulations as well. An interesting
recent development has been proposed in \cite{Introini2014} in the
context of finite-difference discretizations. Their idea is to modify
the fractional step projection scheme such that the Neumann boundary
condition, as introduced in \cite{Kadoch_etal_2012}, appears in the
pressure Poisson equation. Another variant, specifically adopted to
impulsively started flow, is the iterative penalization method proposed
in \cite{Hejlesen2015}. For reviews on immersed boundary and penalization
techniques we refer to \cite{Mittal2005,Peskin2002,Schneider2015}.

The remainder of this article is organized as follows. First we discuss
in section \ref{sec:Numerical-method} the numerical method of \texttt{FluSI}'s
fluid module, which is based on a spectral discretization of the three-dimensional
penalized Navier--Stokes equations. Section \ref{sec:Virtual-insect-model}
describes the module that creates the insect geometry, as well as
the governing equations for free flight simulations, and we focus
on the parallel implementation in section \ref{sec:Parallel-implementation}.
Thereafter we present a validation of our software solution environment
in section \ref{sec:Validation-Tests} for different test cases established
in the literature, and give an outlook for applications to flapping
flight in the turbulent regime in section \ref{sec:Application-to-a}.
Conclusions are drawn in section \ref{sec:Conclusions-and-Perspectives}.

\section{Numerical method\label{sec:Numerical-method}}

\subsection{Model equations}

For the numerical simulation of many flapping flyers, like insects,
the fluid can typically be approximated as incompressible. It is thus
governed by the incompressible Navier--Stokes equations with the no--slip
boundary condition on the fluid--solid interface. The numerical solution
of this problem poses two major challenges. First, the pressure is
a Lagrangian multiplier that ensures the divergence-free condition.
Traditional numerical schemes employ a fractional step or projection
method \cite{Kim1985}, requiring to solve a Poisson problem for the
pressure. Second, the no-slip boundary condition must be satisfied
on a possibly complicated and moving fluid--solid interface, for which
boundary-fitted grids are classical approaches \cite{Thompson1985}.
Since the generation of these grids may be challenging, alternatives
have been developed. The principal idea is to extend the computational
domain to the interior of obstacles. The boundary is then taken into
account by adding supplementary terms to the Navier--Stokes equation.
The technique chosen here is the volume penalization method, which
is physically motivated by replacing the solid obstacle by a porous
medium with small permeability $C_{\eta}$. The penalized version
of the Navier--Stokes equations then reads
\begin{eqnarray}
\partial_{t}\underline{u}+\underline{\omega}\times\underline{u} & = & -\nabla\Pi+\nu\nabla^{2}\underline{u}-\underbrace{\frac{\chi}{C_{\eta}}\left(\underline{u}-\underline{u}_{s}\right)}_{\text{penalization}}-\underbrace{\frac{1}{C_{\mathrm{sp}}}\nabla\times\frac{\left(\chi_{\mathrm{sp}}\underline{\omega}\right)}{\nabla^{2}}}_{\text{sponge}}\label{eq:PNST_org_momentum}\\
\nabla\cdot\underline{u} & = & 0\label{eq:PNST_divergence_free}\\
\underline{u}\left(\underline{x},t=0\right) & = & \underline{u}_{0}\left(\underline{x}\right)\qquad\underline{x}\in\Omega,t>0,\label{eq:PNST_inicond}
\end{eqnarray}
where $\underline{u}$ is the fluid velocity, $\underline{\omega}=\nabla\times\underline{u}$
is the vorticity and $\nu$ is the kinematic viscosity. The sponge
term is explained in section \ref{sub:Wake-removal-techniques}. Eqns.
(\ref{eq:PNST_org_momentum}-\ref{eq:PNST_inicond}) are written in
dimensionless form and the fluid density $\varrho_{f}$ is normalized
to unity. The nonlinear term in eqn. (\ref{eq:PNST_org_momentum})
is written in the rotational form, hence one is left with the gradient
of the total pressure $\Pi=p+\frac{1}{2}\underline{u}\cdot\underline{u}$
instead of the static pressure $p$ \cite{Peyret2002}. This formulation
is chosen because of its favorable properties when discretized with
spectral methods, namely conservation of momentum and energy \cite[pp. 210]{Peyret2002}.
At the exterior of the computational domain, which is supposed to
be sufficiently large, periodic boundary conditions can be assumed.
The mask function $\chi$ is defined as
\begin{equation}
\chi\left(\underline{x},t\right)=\left\{ \begin{array}{cc}
0 & \text{if }\underline{x}\in\Omega_{f}\\
1 & \text{if }\underline{x}\in\Omega_{s}
\end{array}\right.,\label{eq:mask_function_discontinuous}
\end{equation}
where $\Omega_{f}$ is the fluid and $\Omega_{s}$ the solid domain.
In anticipation of the application to moving boundaries, we note that
we will replace the discontinuous $\chi$-function by a smoothed one,
with a thin smoothing layer centered around the interface. The mask
function encodes all geometric information of the problem, and eqns.
(\ref{eq:PNST_org_momentum}-\ref{eq:PNST_inicond}) do not include
no--slip boundary conditions. 

Note that in the fluid domain $\Omega_{f}$ one recovers the original
equation as the penalization term $\frac{\chi}{C_{\eta}}\left(\underline{u}-\underline{u}_{s}\right)$
vanishes. The convergence proof in \cite{Carbou2003,Angot1999} shows
that the solution of the penalized Navier\textendash Stokes equations
(\ref{eq:PNST_org_momentum}-\ref{eq:PNST_inicond}) tends for $C_{\eta}\rightarrow0$
indeed towards the exact solution of Navier\textendash Stokes imposing
no-slip boundary conditions, with a convergence rate of $\mathcal{O}\left(\sqrt{C_{\eta}}\right)$
in the $L^{2}$--norm. The parameter $C_{\eta}$ should thus be chosen
to a small enough value, which is also intuitively clear by the physical
interpretation of $C_{\eta}$ as permeability. However, the choice
of $C_{\eta}$ is subject to constraints, as the penalized equations
are discretized and solved numerically. The modeling error of order
$\mathcal{O}\left(\sqrt{C_{\eta}}\right)$ should be of the same order
as the discretization error \cite{Romain2012}. It is first noted
that in eqn. (\ref{eq:PNST_org_momentum}), $C_{\eta}$ has the dimension
of a time. It is instructive to put the nonlinear, viscous and pressure
terms aside for a moment. One is then left with $\partial_{t}u=-u/C_{\eta}$
inside the solid, with the obvious solution $u=u_{0}\exp\left(-t/C_{\eta}\right)$.
Thus, $C_{\eta}$ can be directly identified as the relaxation time.
Interfering with the time step $\Delta t$, usually implying $\Delta t=\mathcal{O}\left(C_{\eta}\right)$,
this simple fact has important consequences for the numerical solution.
It indicates that a good choice for $C{}_{\eta}$ is not only ``small
enough'', but also ``as large as possible''. Further insight in
the properties of the penalization method can be obtained considering
the penalization boundary layer of width $\delta_{\eta}=\sqrt{\nu C_{\eta}}$,
which forms inside the obstacle $\Omega_{s}$, as shown in \cite{Carbou2003}.
When increasing the resolution, the number of points per boundary
layer thickness, $K=\delta_{\eta}/\Delta x$, should be kept constant,
which implies the relation $C_{\eta}\propto\left(\Delta x\right)^{2}$,
consistent with \cite{Romain2012}, where the penalized Laplace and
Stokes operators were analyzed analytically and numerically. With
the scaling for $C_{\eta}$, one still has to choose the constant
$K$. In fact, for any value of $K$, the method will converge with
the same convergence rate, but the error offset can be tuned. For
the two-dimensional Couette flow, in \cite{Engels2014} we reported
the optimal value of $K=0.128$. In a range of numerical validation
tests, including the ones presented here, we find $K=0.1-0.4$ as
a good choice which we recommend as a guideline for practical applications.

To satisfy the incompressibility constraint (\ref{eq:PNST_divergence_free}),
a Poisson equation for the pressure is derived by taking the divergence
of eqn. (\ref{eq:PNST_org_momentum}), yielding 
\begin{eqnarray}
\nabla^{2}\Pi & = & \nabla\cdot\left(-\underline{\omega}\times\underline{u}-\frac{\chi}{C_{\eta}}\left(\underline{u}-\underline{u}_{s}\right)\right).\label{eq:PNST_UP_pressure_poisson_eqn}
\end{eqnarray}
By construction of the method, eqn. (\ref{eq:PNST_UP_pressure_poisson_eqn})
can be solved with periodic boundary conditions.

The aerodynamic force $\underline{F}$ and the torque moment $\underline{m}$,
both important quantities in the present applications, can be computed
from 
\begin{eqnarray}
\underline{F} & = & \oint_{\partial\Omega_{s}}\sigma\cdot\underline{n}\,\mathrm{d}\gamma=\frac{1}{C_{\eta}}\int_{\Omega}\chi\left(\underline{u}-\underline{u}_{s}\right)\mathrm{d}V+\frac{\mathrm{d}}{\mathrm{d}t}\int_{\Omega_{s}}\underline{u}_{s}\mathrm{d}V\label{eq:force_with_unst_corrections}\\
\underline{m} & = & \oint_{\partial\Omega_{s}}\underline{r}\times\left(\sigma\cdot\underline{n}\right)\,\mathrm{d}\gamma=\frac{1}{C_{\eta}}\int_{\Omega}\underline{r}\times\chi\left(\underline{u}-\underline{u}_{s}\right)\mathrm{d}V+\frac{\mathrm{d}}{\mathrm{d}t}\int_{\Omega_{s}}\underline{r}\times\underline{u}_{s}\mathrm{d}V,\label{eq:moment_of_love}
\end{eqnarray}
where the additional terms are denoted `unsteady corrections' \cite{Uhlmann2005}.

\subsection{Penalization method for moving boundary problems}

The volume penalization method as discussed so far assumes a discontinuous
mask function in the form of eqn. (\ref{eq:mask_function_discontinuous}).
When applying the method to a non-grid aligned body, for example a
circular cylinder, the mask function geometrically approximates the
boundary to first order in $\Delta x$. Results for stationary cylinder
using the discontinuous mask function are acceptable \cite{Schneider2005,Schneider2005a},
but spurious oscillations in the case of moving ones are reported
\cite{Kolomenskiy2009}. The reason is that the discontinuous mask
can be translated only by integer multiples of the grid spacing, and
this jerky motion causes large oscillations in the aerodynamic forces.
Kolomenskiy and Schneider \cite{Kolomenskiy2009} proposed an algorithm
to shift the mask function in Fourier space instead of physical space.
In the present work, we employ a different approach, because displacing
the mask in Fourier space involves (additional) Fourier transforms,
which are computationally expensive, especially if more than one rigid
body is considered, as it is the case here. The idea on hand is to
directly assume the mask function to be smoothed over a thin smoothing
layer \cite{Engels2012a,Engels2014}. To this end, we introduce the
signed distance function $\delta(\underline{x},t)$ \cite{Osher2003},
and the mask function can then be computed from the signed distance,
using 
\begin{equation}
\chi\left(\delta\right)=\left\{ \begin{array}{cc}
1 & \delta\leq-h\\
\frac{1}{2}\left(1+\cos\left(\pi\frac{\delta+h}{2h}\right)\right) & -h<\delta<+h\\
0 & \delta>+h
\end{array}\right.\label{eq:smoothing_mask_as_function_of_delta}
\end{equation}
where the semi-thickness of the smoothing layer, $h$, is used. It
is typically defined relative to the grid size, $h=C_{\mathrm{smth}}\Delta x$,
thus eqn. (\ref{eq:smoothing_mask_as_function_of_delta}) converges
to a Heaviside step function as $\Delta x\rightarrow0$. Nonetheless,
it can be translated by less than one grid point, and then be resampled
on the Eulerian fluid grid.

\subsection{Wake removal techniques\label{sub:Wake-removal-techniques}}

The penalized Navier--Stokes equations (\ref{eq:PNST_org_momentum}-\ref{eq:PNST_inicond})
do, by principle, not include no-slip boundary conditions and furthermore
we discretize them in a periodic domain $\Omega$. In such a periodic
setting, the wake re-enters the domain, which is an undesired artifact.
To overcome it, a supplementary ``sponge'' penalization term can
be added to the vorticity-velocity formulation of the Navier-Stokes
equations, in order to gradually damp the vorticity \cite{Engels2012a,Engels2014}.
The sponge penalization parameter $C_{\mathrm{sp}}$ is usually set
to a larger value than $C_{\eta}$, typically $C_{\mathrm{sp}}=10^{-1}$.
The larger value and its longer relaxation time ensure that if a traveling
vortex pair enters the sponge region, the leading one is not dissipated
too fast, because it otherwise could leave the partner orphaned in
the domain. Applying the Biot-Savart operator to the vorticity-velocity
formulation we formally find $-\frac{1}{C_{\mathrm{sp}}}\nabla\times\frac{\left(\chi_{\mathrm{sp}}\underline{\omega}\right)}{\nabla^{2}}$.
By construction the sponge term is divergence-free, which is important
since otherwise it would contribute to the pressure, which in turn
would be modified even in regions far away from the sponge due to
its nonlocality. Moreover, it leaves the mean flow, i.e., the zeroth
Fourier mode, unchanged. This technique is well adapted to spectral
discretizations, since computing the sponge term requires the solution
of three Poisson problems, which becomes a simple division in Fourier
space. A discussion on the influence of the vorticity sponge can be
found in section \ref{sec:Application-to-a} and Fig. \ref{fig:Influence-sponge}.

Dirichlet conditions on the velocity can be imposed directly with
the volume penalization, which applies for example to channel walls.
For simulations in a uniform, unbounded free-stream, we use both techniques;
in a small layer at the domain borders, the Dirichlet condition $\underline{u}=\underline{u}_{\infty}$
is imposed with the same precision as the actual obstacle, and a preceding,
thicker sponge layer ensures that the upstream influence is minimized
\cite{Engels2014}. The sponge technique is similar to the ``fringe
regions'' proposed in \cite{Schlatter2005}. However, in \cite{Schlatter2005}
only a velocity sponge has been used, which corresponds to the penalization
term, i.e., $-\chi_{\mathrm{sp}}\left(\underline{u}-\underline{u}_{\mathrm{sp}}\right)/C_{\mathrm{sp}}$.
The vorticity sponge idea has the advantage of not requiring a ``desired''
velocity field $\underline{u}_{\mathrm{sp}}$, which has to be set
a priori. It also does not contribute to the pressure, which helps
reducing the region of influence.

\subsection{Discretization}

The model equations (\ref{eq:PNST_org_momentum}) and (\ref{eq:PNST_UP_pressure_poisson_eqn})
can be discretized with any numerical scheme, in particular a Fourier
pseudo-spectral discretization can be used \cite{Schneider2005,Kolomenskiy2009,Engels2014}.
The general idea is to represent field variables as truncated Fourier
series, thus in three dimensions we have for any quantity $\varphi$
(velocity, pressure, vorticity) the discrete complex Fourier coefficients
$\widehat{\varphi}$. They can be computed efficiently with the fast
Fourier transform (\texttt{FFT}) \cite{Cooley1965}. The gradient
of a scalar $q$ can be obtained by multiplying the Fourier coefficients
$\widehat{q}$ with the wavevector $\underline{k}=\left(k_{x},k_{y},k_{z}\right)^{T}$
and the complex unit, $\widehat{\nabla q}=\iota\underline{k}\widehat{q}$.
The Laplace operator becomes a simple multiplication by $-\left|\underline{k}\right|^{2}$.
When using, e.g., finite differences, the dominant part of computational
efforts is spent on solving the Poisson equation in every time step
\cite{Ji2012}. This is a strong motivation to employ a Fourier discretization,
as inverting a diagonal operator becomes a simple division. Inserting
the truncated Fourier series into the model equations and requiring
that the residual vanishes with respect to all test functions (which
are identical with the trial functions $\exp\left(\iota\underline{k}\cdot\underline{x}\right)$
) yields a Galerkin projection and results in an evolution equation
for the Fourier coefficients of the velocity. The nonlinear and penalization
terms contain products, which become convolutions in Fourier space.
To speed-up computation, the products are calculated in physical space.
This last introduces aliasing errors which are virtually eliminated
by the 2/3 rule \cite{Zang1986}, meaning that only 2/3 of the Fourier
coefficients are retained. Such a mixture of spectral and physical
computations is generally labeled \textquotedbl{}pseudo-spectral\textquotedbl{}
and is, when de-aliased, equivalent to a Fourier-Galerkin scheme.
The code can be run with and without dealiasing, but our choice is
to conservatively always turn dealiasing on. More details on the effect
of dealiasing are discussed in \cite[p. 318]{Romain2012}.

The spatially discretized equations can then be advanced in time in
Fourier space. The code provides a classical fourth order Runge--Kutta
with explicit treatment of the diffusion term, and a second order
Adams--Bashforth scheme with integrating factor \cite{Schneider2005}.

Besides the fast solution of Poisson problems, the spectral method
has the advantage of not adding numerical diffusion or dispersion
to the penalized equation, unlike it is the case when discretized
with finite differences. Furthermore, most of the computational effort
is concentrated in the Fourier transforms, which is advantageous from
a computational point of view. However, the discretization requires
the use of an equidistant grid, which implies a large number of grid
points to resolve the thin boundary layers.

Note that we do not explicitly apply a Neumann-type boundary condition
for the pressure in our method. We only use periodic boundary conditions
in the computational domain, which are natural for Fourier methods.
It was shown in \cite{Angot1999} that the pressure field of the penalized
problem converges to the pressure in the original boundary value problem.
The discretization that we use is consistent and stable, therefore
the numerical solution converges to the exact solution of the continuous
penalized problem.

\section{Virtual insect model\label{sec:Virtual-insect-model}}

We previously described the fluid module where the geometry is taken
into account by the penalization method, which is the interface with
the insect module described next.

Insects fly by flapping their wings, which are basically flat with
sharp edges and operate typically at high angle of attack. In the
following, we describe the insect framework used in this work in detail,
the essential task being to construct $\chi$ and $\underline{u}_{s}$
which enter eqn. (\ref{eq:PNST_org_momentum}). The virtual insect
consists of a body and two wings, all of which are performing solid
body rotations around three axes and assumed to be rigid. Therefore,
we will make use of the rotation matrices
\begin{eqnarray*}
R_{x}\left(\xi\right)=\left(\begin{array}{ccc}
1 & 0 & 0\\
0 & \cos\xi & \sin\xi\\
0 & -\sin\xi & \cos\xi
\end{array}\right) & \quad & R_{y}\left(\xi\right)=\left(\begin{array}{ccc}
\cos\xi & 0 & -\sin\xi\\
0 & 1 & 0\\
\sin\xi & 0 & \cos\xi
\end{array}\right)\\
R_{z}\left(\xi\right)=\left(\begin{array}{ccc}
\cos\xi & \sin\xi & 0\\
-\sin\xi & \cos\xi & 0\\
0 & 0 & 1
\end{array}\right)
\end{eqnarray*}
and define the different reference frames, namely the global $\underline{x}^{(g)}$,
body $\underline{x}^{(b)}$, stroke plane $\underline{x}^{(s)}$ and
wing $\underline{x}^{(w)}$, in which the geometry is defined. As
described above, the mask function is constructed in each evaluation
of the right hand side as a function of the signed distance function,
$\chi\left(\underline{x}\right)=\chi\left(\delta\left(\underline{x}\right)\right)$,
according to eqn. (\ref{eq:smoothing_mask_as_function_of_delta}).
\begin{figure}
\begin{centering}
\includegraphics[width=0.75\textwidth]{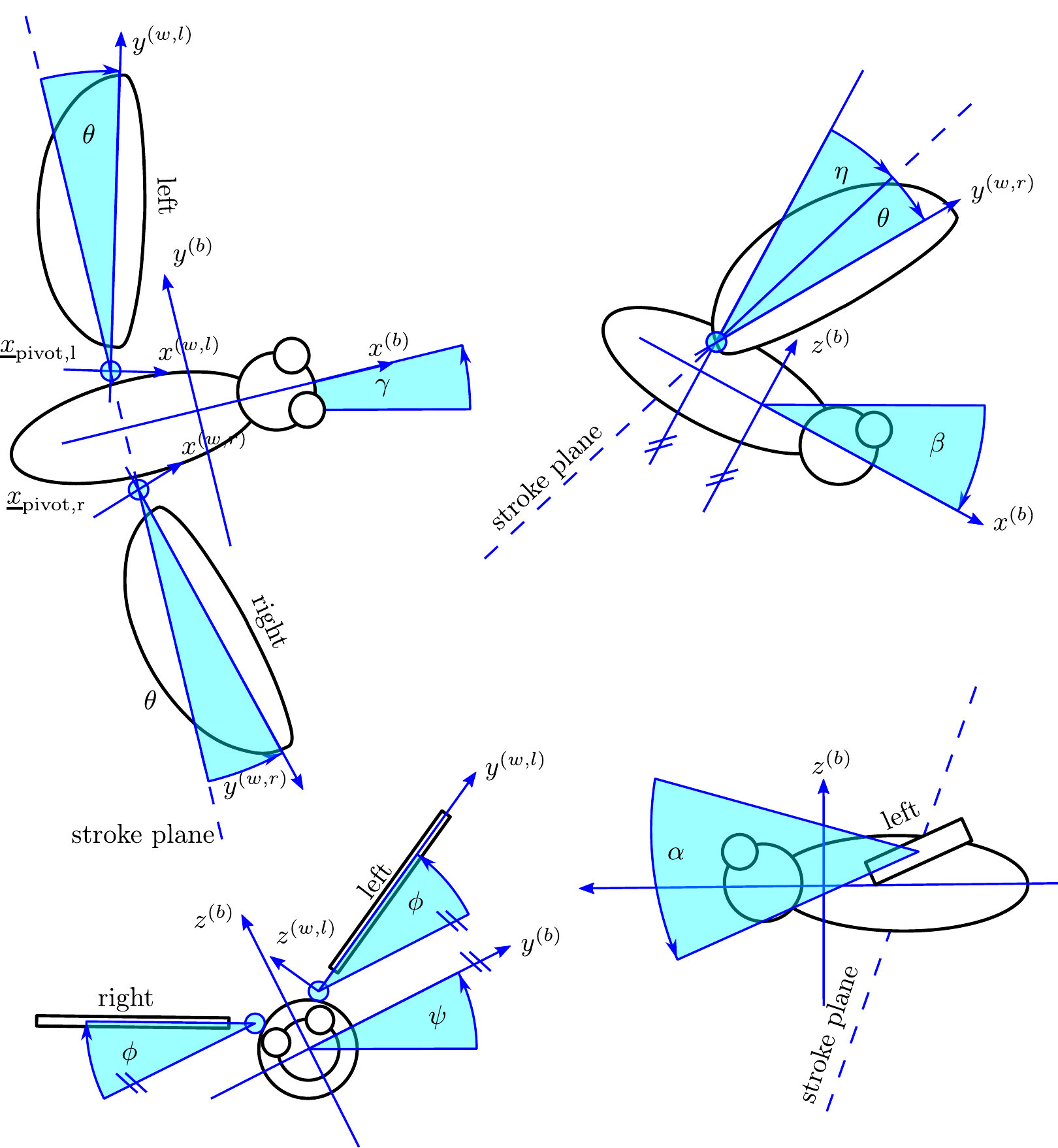}
\par\end{centering}

\caption{Model insect with definitions of the body angles $\gamma$ (yaw),
$\beta$ (pitch), $\psi$ (roll), the anatomical stroke plane angle
$\eta$, the wing coordinate systems and the wing angles $\theta$
(deviation), $\phi$ (position) and $\alpha$ (feathering). All angles
are shown with positive sign. \label{fig:Insect_jerry_definitions}}
\end{figure}

\subsection{Body system}

The insect's body is responsible for a major part of the total drag
force. It is described by its logical center $\underline{x}_{\mathrm{cntr}}^{(g)}$,
the translational velocity $\underline{u}_{\mathrm{cntr}}^{(g)}$
and the body angles $\beta$ (pitch), $\gamma$ (yaw) and $\psi$
(roll), see figure \ref{fig:Insect_jerry_definitions} The center
point $\underline{x}_{\mathrm{cntr}}^{(g)}$ does not necessarily
coincide with the center of gravity, it is rather an arbitrary point
of reference. A point $\underline{x}^{(g)}$ in the global coordinate
system can be transformed to the body system using the following linear
transformation
\begin{eqnarray}
\underline{x}^{(b)} & = & M_{\mathrm{body}}\left(\psi,\beta,\gamma\right)\left(\underline{x}^{(g)}-\underline{x}_{\mathrm{cntr}}^{(g)}\right)\nonumber \\
M_{\mathrm{body}} & = & R_{x}\left(\psi\right)R_{y}\left(\beta\right)R_{z}\left(\gamma\right).\label{eq:M_body_yawpitchroll}
\end{eqnarray}
Since rotation matrices do not commute, it is important to note that
the body is first yawed, then pitched and finally rolled, which is
conventional in flight mechanics. The geometry of the body is defined
in the body reference frame. The angular velocity of the body in the
global system is
\begin{eqnarray*}
\underline{\Omega}_{b}^{(g)} & = & R_{z}^{-1}\left(\gamma\right)\left[\left(\begin{array}{c}
0\\
0\\
\dot{\gamma}
\end{array}\right)+R_{y}^{-1}\left(\beta\right)\left[\left(\begin{array}{c}
0\\
\dot{\beta}\\
0
\end{array}\right)+R_{x}^{-1}\left(\psi\right)\left(\begin{array}{c}
\dot{\psi}\\
0\\
0
\end{array}\right)\right]\right]\\
\underline{\Omega}_{b}^{(b)} & = & M_{\mathrm{body}}\:\Omega_{b}^{(g)}
\end{eqnarray*}
which defines the velocity field inside the insect resulting from
the body motion,
\begin{equation}
\underline{u}_{b}^{(g)}=\underline{u}_{\mathrm{cntr}}^{(g)}+M_{\mathrm{body}}^{-1}\left(\underline{\Omega}_{b}^{(b)}\times\underline{x}^{(b)}\right).\label{eq:body_velocity_field}
\end{equation}
Equation (\ref{eq:body_velocity_field}) is valid also in the wings,
since the flapping motion is prescribed relative to the body.

\subsection{Body shape}

The body shape is described in the body reference frame described
previously. For instance, for the body depicted in figure \ref{fig:Insect_jerry_definitions},
which is composed of an ellipsoidal shaped thorax and spheres for
the head and eyes, the signed distance function is the intersection
of the distance functions for the thorax, head and eyes,
\begin{equation}
\delta_{\mathrm{body}}=\max\left(\delta_{\mathrm{thorax}},\delta_{\mathrm{head}},\delta_{\mathrm{eyes}}\right).\label{eq:DELTA_body_jerry}
\end{equation}
The $\max$ operator of the signed distances in eqn. (\ref{eq:DELTA_body_jerry})
represents the intersection operator \cite{Osher2003}. The signed
distances for the components read
\begin{eqnarray*}
\delta_{\mathrm{thorax}}\left(\underline{x}^{(b)}\right) & = & \sqrt{\left(y^{(b)}\right)^{2}+\left(z^{(b)}\right)^{2}}-\sqrt{b^{2}\left(1-\left(x^{(b)}/a\right)^{2}\right)}\\
\delta_{\mathrm{head}}\left(\underline{x}^{(b)}\right) & = & \left|\underline{x}^{(b)}-\underline{x}_{0,\mathrm{head}}^{(b)}\right|\\
\delta_{\mathrm{eyes}}\left(\underline{x}^{(b)}\right) & = & \left|\underline{x}^{(b)}-\underline{x}_{0,\mathrm{eyes}}^{(b)}\right|,
\end{eqnarray*}
where $a$, $b$ define the axes of the thorax ellipsoid and $\underline{x}_{0,\mathrm{head,eyes}}^{(b)}$
are the centers of the spheres.

\subsection{Wing system}

We consider only insects with two wings, one on each side, that are
rotating about the pivot points $\underline{x}_{\mathrm{pivot,r}}^{(b)}$
and $\underline{x}_{\mathrm{pivot,l}}^{(b)}$. These pivots do not
necessarily lie on the body surface; we rather allow a gap between
wings and body. This gap avoids problems with non-solenoidal velocity
fields at the wing base. It is conventional to introduce a stroke
plane, which is a plane tilted with respect to the body by an angle
$\eta$. The coordinate in the stroke plane reads
\[
\underline{x}^{(s)}=M_{\mathrm{stroke}}\left(\underline{x}^{(b)}-\underline{x}_{\mathrm{pivot}}^{(b)}\right).
\]
Here, we use an anatomical stroke angle, that is, the angle $\eta$
is defined relative to the body. Within the stroke plane, the wing
motion is described by the angles $\alpha$ (feathering angle or angle
of attack), $\phi$ (positional or flapping angle), $\theta$ (deviation
or out-of-stroke angle). Applying two rotation matrices yields the
transformation from the body to the wing coordinate system: 
\[
\underline{x}^{(w)}=M_{\mathrm{wing}}\underline{x}^{(s)}=M_{\mathrm{wing}}M_{\mathrm{stroke}}\left(\underline{x}^{(b)}-\underline{x}_{\mathrm{pivot}}^{(b)}\right).
\]
When flapping symmetrically, i.e., both wings following the same motion
protocol, the stroke and wing rotation matrices for the left and right
wing are given by
\begin{eqnarray*}
M_{\mathrm{stroke,l}}=R_{y}\left(\eta\right) & \qquad & M_{\mathrm{stroke,r}}=R_{x}\left(\pi\right)R_{y}\left(\eta\right)\\
M_{\mathrm{wing,l}}=R_{y}\left(\alpha\right)R_{z}\left(-\theta\right)R_{x}\left(\phi\right) & \qquad & M_{\mathrm{wing,r}}=R_{y}\left(-\alpha\right)R_{z}\left(-\theta\right)R_{x}\left(-\phi\right)
\end{eqnarray*}
due to the rotation $R_{x}\left(\pi\right)$ the sign of $\theta$
for the right wing does not have to be inverted. The angular velocities
of the wings are given by 
\begin{eqnarray*}
\underline{\Omega}_{w}^{(b)} & = & M_{\mathrm{stroke}}^{-1}\left[R_{x}^{-1}\left(\phi\right)\left[\left(\begin{array}{c}
\dot{\phi}\\
0\\
0
\end{array}\right)+R_{z}^{-1}\left(-\theta\right)\left[\left(\begin{array}{c}
0\\
0\\
-\dot{\theta}
\end{array}\right)+R_{y}^{-1}\left(\alpha\right)\left(\begin{array}{c}
0\\
\dot{\alpha}\\
0
\end{array}\right)\right]\right]\right]\\
\underline{\Omega}_{w}^{(w)} & = & M_{\mathrm{wing}}\:\underline{\Omega}_{w}^{(b)}
\end{eqnarray*}
which is used to compute the velocity field resulting from the wing
motion, 
\begin{eqnarray}
\underline{u}_{w}^{(w)} & = & \underline{\Omega}_{w}^{(w)}\times\underline{x}^{(w)}\nonumber \\
\underline{u}_{w}^{(g)} & = & M_{\mathrm{body}}^{-1}\;M_{\mathrm{wing}}^{-1}\;M_{\mathrm{stroke}}^{-1}\;\underline{u}_{w}^{(w)}.\label{eq:wing_velocity_field}
\end{eqnarray}
The total velocity field inside the wings is given as the superposition
of the body and wing rotation,
\[
\underline{u}_{s}^{(g)}\left(\underline{x}\in\left\{ \underline{x}_{w}\right\} \right)=\underline{u}_{w}^{(g)}+\underline{u}_{b}^{(g)}.
\]
The actual kinematics, i.e., the angles $\alpha\left(t\right)$, $\phi\left(t\right)$
and $\theta\left(t\right)$ are parametrized by either Fourier or
Hermite interpolation coefficients and read from a \texttt{{*}.ini}
file.

\subsection{Wing shape}

\begin{figure}
\begin{centering}
\includegraphics[scale=0.75]{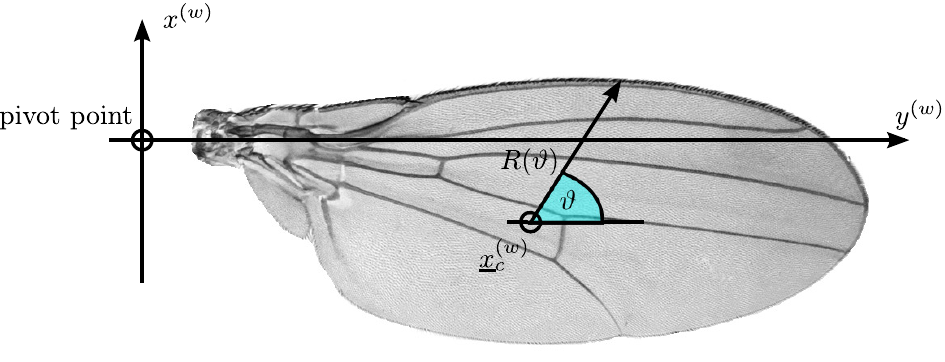}
\par\end{centering}

\caption{Realistic wing shapes are described in polar coordinates $R\left(\vartheta\right)$.\label{fig:Realistic-wing-shapes-discreti}}
\end{figure}

In the previous section we defined the wing reference frame $\underline{x}^{(w)}$,
in which we now describe the wing's signed distance function $\delta$.
In general, we define a set of several signed distance functions,
each of which describes one surface of the wing. The signed distance
function of the entire wing is then given by their intersection. For
some model insects, we consider simple wings for which straightforward
analytical expressions are available. For a rectangular wing, for
instance the one illustrated in figure \ref{fig:suzuki_ALL} (c),
we find for the signed distance function 
\begin{equation}
\delta\left(\underline{x}^{(w)}\right)=\max\left(x^{(w)}-b;x^{(w)}-\left(B-b\right);y^{(w)}-1;a-y^{(w)};\left|z\right|-h/2\right).\label{eq:signed_distance_rectangular_wing}
\end{equation}
For realistic insect wings however, we parametrize the wing shape
in polar coordinates. As illustrated in figure \ref{fig:Realistic-wing-shapes-discreti},
the shape in the wing plane is described by the center point $\underline{x}_{c}^{(w)}$,
which is arbitrary as long as the function $R\left(\vartheta\right)$
is unique for all $\vartheta$. To sample the wing on a computational
grid, we need a function $R\left(\vartheta\right)$ that can be evaluated
for all $\vartheta$. As $R\left(\vartheta\right)$ is naturally $2\pi$-periodic,
a truncated Fourier series can be used:
\begin{equation}
R\left(\vartheta\right)=\frac{a_{0}}{2}+\sum_{i=1}^{N}a_{i}\cos\left(2\pi i\vartheta\right)+\sum_{i=1}^{N}b_{i}\sin\left(2\pi i\vartheta\right)\label{eq:Fourier_radious_wing}
\end{equation}
In practice, eqn (\ref{eq:Fourier_radious_wing}) has to be evaluated
for all grid points in the vicinity of the wing, which requires $\mathcal{O}\left(N_{x}N_{y}N_{z}\right)$
evaluations with a small constant. The computational cost can however
be significant, which is why eqn. (\ref{eq:Fourier_radious_wing})
is evaluated for 25 000 values of $\vartheta$ once during initialization.
Afterwards, linear interpolation is used for its lower computational
cost. The signed distance for such a wing then reads
\[
\delta\left(\underline{x}^{(w)}\right)=\max\left(\left|z^{(w)}\right|-h/2;r\left(\vartheta\right)-R\left(\vartheta\right)\right).
\]
If a wing cannot be described by one radius, because $R\left(\vartheta\right)$
is not unique, several radii and center points can be used \cite{Bimbard2013}.

\subsection{Power requirement}

Actuating the wings requires power expenditures that are very difficult
to measure directly. In numerical simulations, the power can be obtained
directly, since the aerodynamic torque moment $\underline{m}$ with
respect to the wing pivot point is available from eqn. (\ref{eq:moment_of_love}).
The power $P_{\mathrm{aero}}$ required to move one wing is found
to be 
\begin{equation}
P_{\mathrm{aero}}=-\underline{m}\cdot\left(\underline{\Omega}_{w}-\underline{\Omega}_{b}\right)\label{eq:insect_power}
\end{equation}
which is equivalent to the definition $P_{\mathrm{aero}}=\int\underline{u}\mathrm{d}\underline{F}$
given in \cite{Maeda2013}. In addition to the aerodynamic power,
the inertial power has to be expended, i.e., the power required to
move the wing in vacuum. As the flapping motion is periodic, its stroke
averaged value is zero. The inertial power $P_{\mathrm{inert}}$ is
positive if the wing is accelerated (power consumed) and negative
if it is decelerated. The definition is
\[
P_{\mathrm{inert}}=\underline{\Omega}_{w}^{(w)}\cdot\left(\underline{\underline{J}}^{(w)}\underline{\dot{\Omega}}_{w}^{(w)}+\underline{\Omega}_{w}^{(w)}\times\underline{\underline{J}}^{(w)}\underline{\Omega}_{w}^{(w)}\right)
\]
with the wing tensor of inertia $\underline{\underline{J}}^{(w)}$
\cite{Berman2007}. The sum of inertial and aerodynamic power can
be negative during deceleration phases, which would mean that the
insect can store energy in its muscles. It is unknown to what extent
this can be realized, and it is thus often conservatively assumed
that energy storage is not possible, in which case the total power
$P_{\mathrm{total}}$ is given by $P_{\mathrm{total}}=\max\left(P_{\mathrm{inert}}+P_{\mathrm{aero}},0\right).$

\subsection{Governing equations in free flight}

Until now we have considered the insect to be fixed, i.e., tethered
in the computational domain. In free flight, additional equations
have to be solved together with the fluid, namely Newton's law. The
body translation is then governed by 
\[
M\,\underline{\dot{u}}_{\mathrm{cntr}}^{(g)}=\underline{F}^{(g)}
\]
where $\underline{F}^{(g)}$ contains the aerodynamic and gravitational
forces and $M$ is the mass of the insect. For simplicity, $\underline{x}_{\mathrm{centr}}$
and $\underline{u}_{\mathrm{centr}}$ correspond to the center of
gravity in the case of free flight. To handle the rotational degrees
of freedom, we employ a quaternion based formulation, similar to the
one proposed in \cite{Maeda2010}, which avoids the `Gimbal lock'
problem. The governing equation for the angular velocity $\underline{\Omega}^{(b)}$
in the body reference frame reads
\[
\underline{\underline{J}}^{(b)}\dot{\underline{\Omega}}^{(b)}+\left(\begin{array}{ccc}
0 & -\Omega_{z}^{(b)} & \Omega_{y}^{(b)}\\
\Omega_{z}^{(b)} & 0 & -\Omega_{x}^{(b)}\\
-\Omega_{y}^{(b)} & \Omega_{x}^{(b)} & 0
\end{array}\right)\underline{\underline{J}}^{(b)}\underline{\Omega}^{(b)}=\underline{m}^{(b)},
\]
where $\underline{\underline{J}}^{(b)}$ is the moment of inertia
around the body axes $\left(x^{(b)},y^{(b)},z^{(b)}\right)$ and $\underline{m}$
is the vector of torque moments as defined in eqn. (\ref{eq:moment_of_love}).
The skew-symmetric term stems from the change into a moving reference
frame. We introduce the normalized quaternion $\varepsilon$ with
components $\varepsilon_{i}$, $i=0,\ldots,3$, $\sum\varepsilon_{i}^{2}=1$.
The governing equations for the quaternion state are 
\[
\dot{\underline{\varepsilon}}=\frac{1}{2}\underline{\underline{S}}^{T}\,\underline{\Omega}^{(b)}
\]
with the matrix
\[
S=\left(\begin{array}{cccc}
-\varepsilon_{1} & \varepsilon_{0} & \varepsilon_{3} & -\varepsilon_{2}\\
-\varepsilon_{2} & -\varepsilon_{3} & \varepsilon_{0} & \varepsilon_{1}\\
-\varepsilon_{3} & \varepsilon_{2} & -\varepsilon_{1} & \varepsilon_{0}
\end{array}\right).
\]
Assuming $\underline{\underline{J}}^{(b)}$ to be constant the body
axes to be the prinicipal axes of inertia (i.e., $\underline{\underline{J}}^{(b)}$
is diagonal), the following first order system set of 13 equations
is obtained 
\begin{eqnarray}
\frac{\mathrm{d}}{\mathrm{d}t}\left(\begin{array}{c}
x_{\mathrm{cntr}}^{(g)}\\
y_{\mathrm{cntr}}^{(g)}\\
z_{\mathrm{cntr}}^{(g)}\\
u_{\mathrm{cntr,x}}^{(g)}\\
u_{\mathrm{cntr,y}}^{(g)}\\
u_{\mathrm{cntr,z}}^{(g)}\\
\varepsilon_{0}\\
\varepsilon_{1}\\
\varepsilon_{2}\\
\varepsilon_{3}\\
\Omega_{x}^{(b)}\\
\Omega_{y}^{(b)}\\
\Omega_{z}^{(b)}
\end{array}\right) & = & \left(\begin{array}{c}
u_{\mathrm{cntr,x}}^{(g)}\\
u_{\mathrm{cntr,y}}^{(g)}\\
u_{\mathrm{cntr,z}}^{(g)}\\
F_{x}^{(g)}/M\\
F_{y}^{(g)}/M\\
\left(F_{z}^{(g)}/M-g\right)\\
\left(-\varepsilon_{1}\Omega_{x}^{(b)}-\varepsilon_{2}\Omega_{y}^{(b)}-\varepsilon_{3}\Omega_{z}^{(b)}\right)/2\\
\left(\varepsilon_{0}\Omega_{x}^{(b)}-\varepsilon_{3}\Omega_{y}^{(b)}+\varepsilon_{2}\Omega_{z}^{(b)}\right)/2\\
\left(\varepsilon_{3}\Omega_{x}^{(b)}+\varepsilon_{0}\Omega_{y}^{(b)}-\varepsilon_{1}\Omega_{z}^{(b)}\right)/2\\
\left(-\varepsilon_{2}\Omega_{x}^{(b)}+\varepsilon_{1}\Omega_{y}^{(b)}+\varepsilon_{0}\Omega_{z}^{(b)}\right)/2\\
\left(\left(J_{y}^{(b)}-J_{z}^{(b)}\right)\Omega_{y}^{(b)}\Omega_{z}^{(b)}+m_{x}^{(b)}\right)/J_{x}^{(b)}\\
\left(\left(J_{z}^{(b)}-J_{x}^{(b)}\right)\Omega_{z}^{(b)}\Omega_{x}^{(b)}+m_{y}^{(b)}\right)/J_{y}^{(b)}\\
\left(\left(J_{x}^{(b)}-J_{y}^{(b)}\right)\Omega_{x}^{(b)}\Omega_{y}^{(b)}+m_{z}^{(b)}\right)/J_{z}^{(b)}
\end{array}\right)\label{eq:oh_my_god_look_a_vector}
\end{eqnarray}
which is solved with the same time discretization as the fluid. The
rotation matrix $M_{body}$ is then computed from the quaternion $\varepsilon_{i}$
\begin{equation}
M_{body}=\left(\begin{array}{ccc}
\varepsilon_{0}^{2}+\varepsilon_{1}^{2}-\varepsilon_{2}^{2}-\varepsilon_{3}^{2} & 2\left(\varepsilon_{1}\varepsilon_{2}-\varepsilon_{3}\varepsilon_{0}\right) & 2\left(\varepsilon_{1}\varepsilon_{3}+\varepsilon_{2}\varepsilon_{0}\right)\\
2\left(\varepsilon_{1}\varepsilon_{2}+\varepsilon_{3}\varepsilon_{0}\right) & \varepsilon_{0}^{2}-\varepsilon_{1}^{2}+\varepsilon_{2}^{2}-\varepsilon_{3}^{2} & 2\left(\varepsilon_{2}\varepsilon_{3}-\varepsilon_{1}\varepsilon_{0}\right)\\
2\left(\varepsilon_{1}\varepsilon_{3}-\varepsilon_{2}\varepsilon_{0}\right) & 2\left(\varepsilon_{2}\varepsilon_{3}+\varepsilon_{1}\varepsilon_{0}\right) & \varepsilon_{0}^{2}-\varepsilon_{1}^{2}-\varepsilon_{2}^{2}+\varepsilon_{3}^{2}
\end{array}\right),\label{eq:}
\end{equation}
which replaces the definition in eqn. (\ref{eq:M_body_yawpitchroll})
in the free-flight case. The initial values at time $t=0$ for $\varepsilon_{i}$
can conveniently be computed from a set of yaw, pitch and roll angles,
\[
\left(\begin{array}{c}
\varepsilon_{0}\\
\varepsilon_{1}\\
\varepsilon_{2}\\
\varepsilon_{3}
\end{array}\right)=\left(\begin{array}{c}
\cos\left(\psi/2\right)\cos\left(\beta/2\right)\cos\left(\gamma/2\right)+\sin\left(\psi/2\right)\sin\left(\beta/2\right)\sin\left(\gamma/2\right)\\
\sin\left(\psi/2\right)\cos\left(\beta/2\right)\cos\left(\gamma/2\right)-\cos\left(\psi/2\right)\sin\left(\beta/2\right)\sin\left(\gamma/2\right)\\
\cos\left(\psi/2\right)\sin\left(\beta/2\right)\cos\left(\gamma/2\right)+\sin\left(\psi/2\right)\cos\left(\beta/2\right)\sin\left(\gamma/2\right)\\
\cos\left(\psi/2\right)\cos\left(\beta/2\right)\sin\left(\gamma/2\right)-\sin\left(\psi/2\right)\sin\left(\beta/2\right)\cos\left(\gamma/2\right)
\end{array}\right).
\]
In the actual implementation, we multiply the right hand side of equation
(\ref{eq:oh_my_god_look_a_vector}) with a six component vector with
zeros or ones, to deactivate some degrees of freedom.

\section{Parallel implementation\label{sec:Parallel-implementation}}

Until now we described the numerical method employed by the \texttt{FluSI}
code, which is based on the volume penalization method combined with
a pseudo-spectral discretization. This framework allows for a high
degree of modularization, since all geometry is encoded in the mask
function and the solid velocity field. The framework is intended to
be applicable also for higher Reynolds number flow, for which small
spatial and temporal vortical structures appear. To resolve these
scales, high resolution and therefore the usage of high-performance
computers is required. To this end, our code is designed to compute
on massively parallel machines with distributed memory architectures.
The parallel implementation is based on the \texttt{MPI} protocol
and written in \texttt{FORTRAN95}. 

\begin{figure}
\centering{}\includegraphics[width=0.5\textwidth]{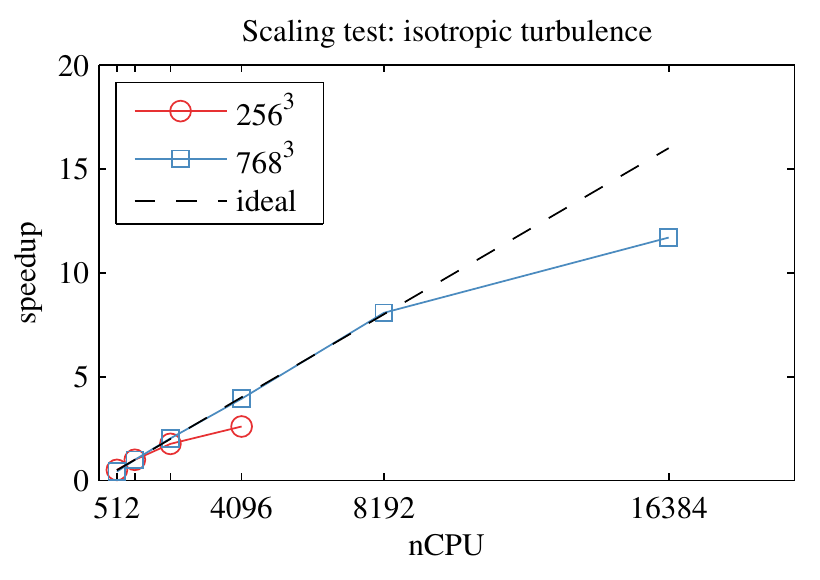}\includegraphics[width=0.5\textwidth]{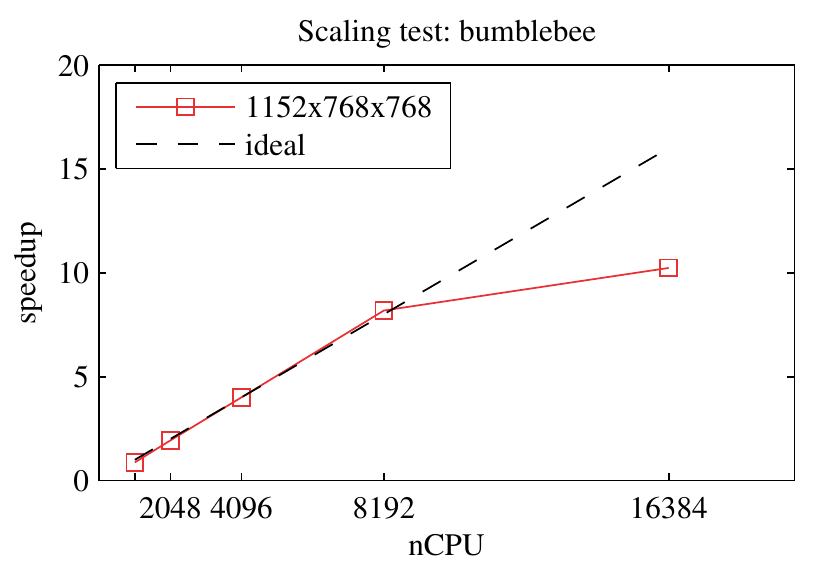}\caption{\label{fig:scaling}Parallel scaling tests on a large-scale computing
cluster of type IBM BlueGene/Q. Left: isotropic turbulence simulation
without insect, for two different resolutions. Fifty time steps have
been performed. The reference simulation is the 1024 run. Right: scaling
test for a bumblebee simulation (see section \pageref{sec:Application-to-a}).
Two complete strokes have been computed, and the timing is averaged
over the second one. The reference simulation is the 2048 cores run. }
\end{figure}

For the fluid part the \texttt{P3DFFT} library is used to compute
the 3D Fourier transforms \cite{Pekurovsky2012}. This library provides
a parallel data decomposition framework, and employs the \texttt{FFTW}
library \cite{Frigo2005} for the one--dimensional Fourier transforms.
The flow variables are stored on the three--dimensional Cartesian
computational grid. Each \texttt{MPI} process only holds a portion
of the total data, and the parallel decomposition is performed on
at most two indices, i.e., a pencil decomposition is used. The $x$-direction
is not split among processes, the code can thus run on $N_{y}N_{z}$
processes at most. This limitation is however not of practical importance,
since $N_{y}N_{z}$ usually exceeds the number of available CPUs by
orders of magnitude. Besides the Fourier transforms, the operations
are local, i.e., they do not require inter-processor communication.
The parallel scaling test for the fluid module is shown in Fig. \ref{fig:scaling}
(left), for the case of an isotropic turbulence simulation and two
different problem sizes, $256^{3}$ and $768^{3}$. The latter shows
good scaling up to 8192 CPU on the IBM BlueGene/Q machine\footnote{http://www.idris.fr/eng/turing/}
used for the tests.

The insect module is used to generate the $\chi$-function and the
solid body velocity field $\underline{u}_{s}$. It is crucial that
the implementation does not spoil the favorable scaling properties.
The module is therefore designed around an derived insect datatype
called \texttt{diptera}, which holds all properties of the insect,
such as the wing shape, body position and Fourier coefficients of
the wing kinematics. This object is of negligible size, and therefore
each MPI process holds a copy of it. The task to construct the penalization
term is then again completely local. Forces and torques (eqn. (\ref{eq:force_with_unst_corrections}-\ref{eq:moment_of_love}))
are computed efficiently with the \texttt{MPI\_allreduce} command.
It is further necessary to distinguish between distinct parts of the
mask function, e.g., body and wings and possibly outer boundary conditions,
like channel walls. To accomplish this, we introduce the concept of
mask coloring, i.e., we allocate a second \texttt{integer{*}2} array
for the mask function that holds the value of the mask color. This
allows to easily exclude for example channel walls from the force
computation in eqn. (\ref{eq:force_with_unst_corrections}), and limits
the memory footprint to two arrays. We further use it to compute the
forces acting on the body and wings individually. If the insect's
body is tethered, there is no need to reconstruct it every time the
mask is updated. The mask coloring can in this case be used easily
to delete only the wings from the mask function while keeping the
body in memory. This reduces the CPU time requirement of the mask
function, since the body occupies a relatively large volume (compared
to the wings) and consequently a larger number of grid points are
affected. 

Fig. \ref{fig:scaling} (right) shows the scaling test of an insect
simulation. The insect model is the bumblebee presented in section
\ref{sec:Application-to-a}, but the scaling behavior is the same
for any other model. We computed two complete strokes at a resolution
of $1152\times768\times768$, and the elapsed walltime was measured
for the second one. The parallel scaling is virtually the same as
in the pure fluid case, indicating that the insect module does indeed
not spoil parallel scaling efficiency. 

\begin{algorithm}
\begin{enumerate}
\item Initialization. Read parameters from \texttt{{*}.ini} file, allocate
memory, set up fluid initial condition, initialize \texttt{P3DFFT}
\item Begin loop over time step

\begin{enumerate}
\item Generate mask function $\chi\left(\underline{x},t^{n}\right)$ and
solid velocity field $\underline{u}_{s}\left(\underline{x},t^{n}\right)$
\item Determine time step $\Delta t$ 
\item Compute sources $\underline{F}\left(\underline{u}^{n}\right)=-\underline{\omega}^{n}\times\underline{u}^{n}-\frac{\chi^{n}}{C_{\eta}}\left(\underline{u}^{n}-\underline{u}_{s}^{n}\right)-\frac{1}{C_{\mathrm{sp}}}\nabla\times\frac{\left(\chi_{\mathrm{sp}}\underline{\omega}^{n}\right)}{\nabla^{2}}$
\item Add pressure gradient $\underline{F}^{n}\leftarrow\underline{F}^{n}-\nabla\left[\left(\nabla\cdot\underline{F}^{n}\right)/\nabla^{2}\right]$
\item Advance fluid to new time level $\underline{u}^{n+1}=\mathrm{AB2}\left(\underline{u}^{n},\underline{F}^{n}\right)$
\item Optional: Advance free-flight equations to new time level, using the
$\mathrm{AB2}$ scheme
\item Output: write flow statistics and flow field data to disk
\end{enumerate}
\item When terminal time is reached, free memory and quit.
\end{enumerate}
\caption{\label{alg:General-process-for-simulations}General process for a
simulation.}
\end{algorithm}

\begin{figure}
\begin{centering}
\includegraphics[width=0.75\textwidth]{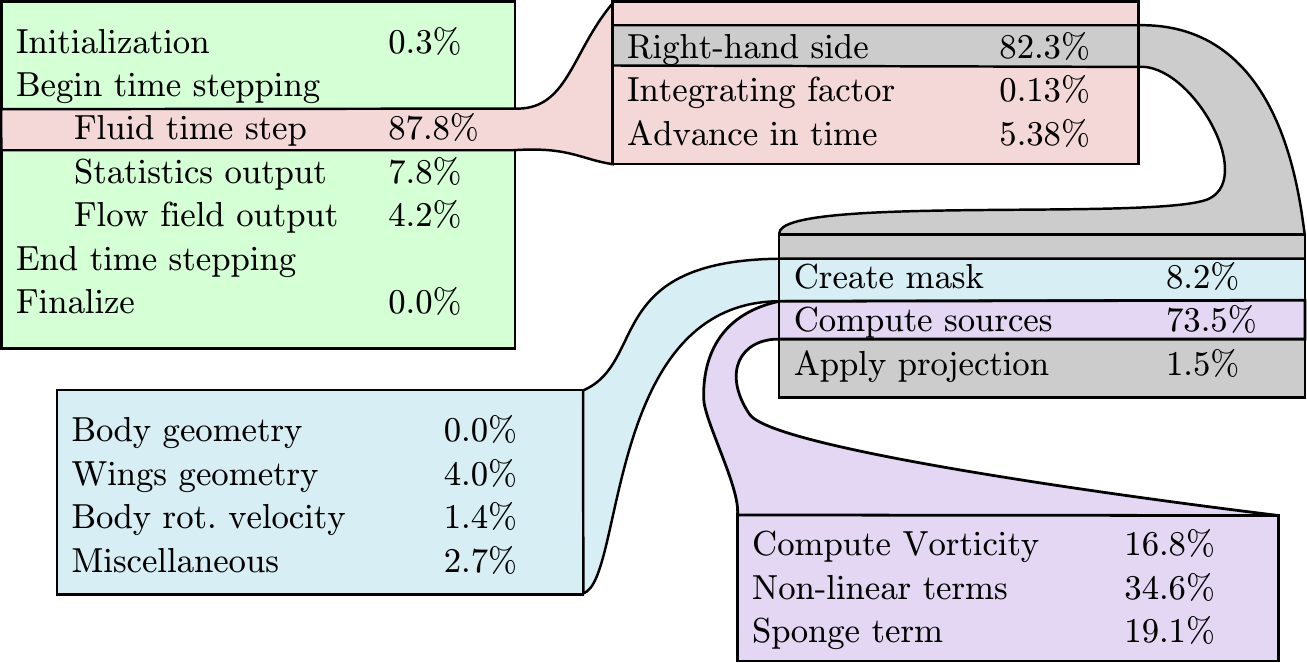}
\par\end{centering}

\caption{\label{fig:timings}Partition of computational efforts spent on a
bumblebee simulation (see section \pageref{sec:Application-to-a}).
The resolution is $1152\times768\times768$. Most efforts are spent
on advancing the fluid in time, which in turn is largely dominated
by the computation of fluid source terms. All per-cent is with respect
to total execution time.}
\end{figure}

The general process for a simulation is summarized in algorithm \ref{alg:General-process-for-simulations}.
All parameters, like the resolution etc., are read from a \texttt{{*}.ini}
file, which avoids recompiling the code every time a parameter is
changed. The code writes time series of quantities like the fluid
forces, enstrophy or aerodynamic power to ascii \texttt{{*}.t} files.
Flow field data, such as the velocity and pressure fields, are stored
using the hierarchical data format (\texttt{HDF5}) library, in order
to guarantee maximum compatibility with third-party applications,
such as the open-source visualization software \texttt{Paraview},
which is used for the visualizations shown in the results section.
Figure \ref{fig:timings} shows how the computing time spent on various
tasks is distributed. The compuation is dominated by the fluid time
stepping (87.8\%); the remaing parts are used in the computation of
flow statistics and field output. Computing the source terms is by
far the most significant contribution, and the generation of the geometry
only consumes about 8\% of the total time.

\section{Validation Tests\label{sec:Validation-Tests}}

For code validation we consider four test cases with increasing complexity,
a falling sphere, a rectangular flapping wing, a hovering flight of
a fruit fly and the free flight of a butterfly model.

\subsection{Falling sphere}

\begin{figure}[t]
\begin{centering}
\includegraphics[scale=0.75]{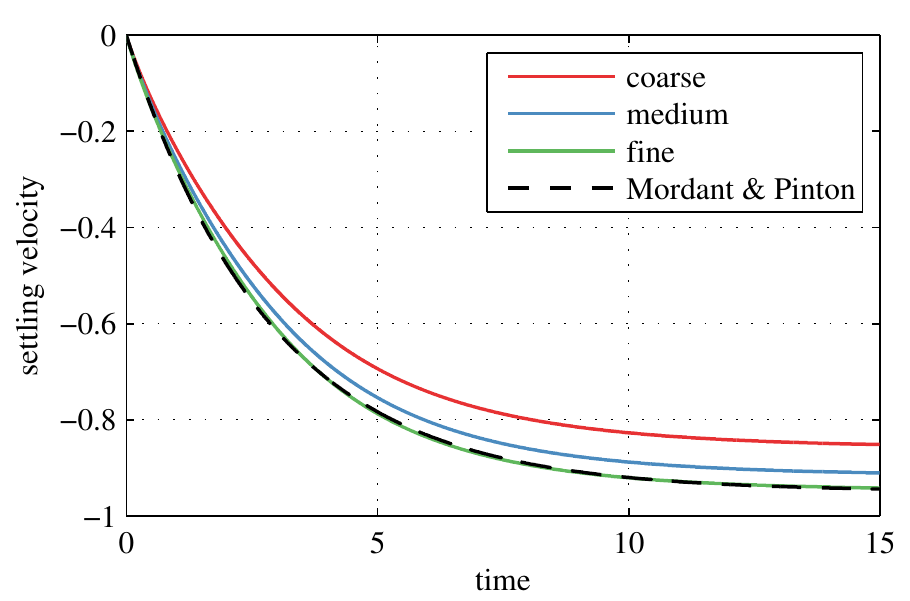}
\par\end{centering}

\caption{Settling velocity of a falling sphere, experimental results by Mordant
\& Pinton \cite{Mordant2000} and present results for coarse, medium
and fine grids.\label{fig:Settling-velocity-of}}

\end{figure}

The first test case to validate the flow solver is the sedimentation
of a sphere, which in our terminology is an insect without wings and
with a spherical body. We consider case 1 proposed by Mordant and
Pinton \cite{Mordant2000}, who studied the sedimentation problem
experimentally in a water tank. The sphere of unit diameter and mass
$M=1.3404$ is falling in fluid of viscosity $\nu=0.0228$ and unit
density. The dimensionless gravity is $g=0.8036$ and the terminal
settling velocity obtained from the experiments is $U=0.9488$. We
perform a grid convergence test using the domain size $8\times8\times16$,
and parameters $N_{x}\times N_{y}\times N_{z}\times C_{\eta}$ of
a coarse ($96\times96\times192\times10^{-3}$), medium ($192\times192\times384\times2.5\cdot10^{-4}$)
and fine ($384\times384\times768\times6.25\cdot10^{-5}$) grid. The
number of points per penalization boundary layer is $K=0.0573$. The
results of the convergence study are illustrated in figure \ref{fig:Settling-velocity-of}
and the settling velocity for the finest resolution differs from the
experimental findings by less than 1\%. The finest resolution required
30 GB of memory and 34 400 CPU hours on 1024 cores to perform 266
667 time steps. The computational cost is relatively high, since small
values of $C_{\eta}$ are required as the Reynolds number is small.

\subsection{Validation case of a rectangular flapping wing}

\begin{figure}[t]
\begin{centering}
\includegraphics[width=1\textwidth]{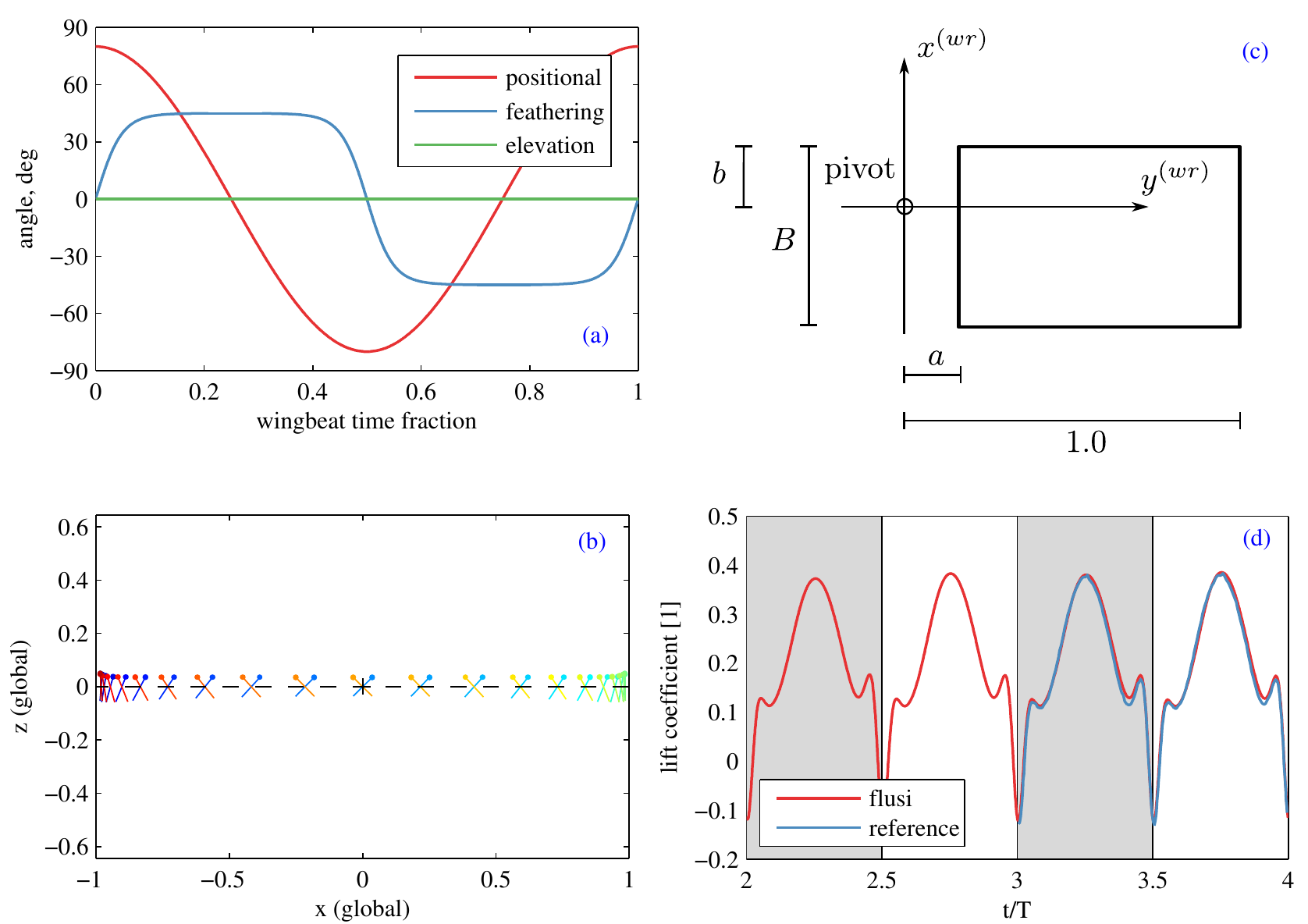}
\par\end{centering}

\caption{Flapping rectangular wing. (a) kinematics used in the test case, as
given by Suzuki et al. \cite{Suzuki2015}. (b) visualization of the
wing kinematics by a chord section (without body, color represents
time) (c) Geometry of the flapping rectangular wing. Contrary to \cite{Suzuki2015},
we normalize the distance pivot-wing tip to unity. (d) time evolution
of the vertical force acting on the wing for the last two strokes,
with the reference solution presented in \cite{Suzuki2015}.\label{fig:suzuki_ALL}}
\end{figure}

We consider the setup proposed by Suzuki and co-workers in \cite{Suzuki2015}
Appendix B2. It considers only one rectangular wing with a finite
thickness, neglecting thus the body and the second wing. The fact
that the thickness is finite and the geometry rather simple, compared
to actual insects, motivates the choice of this setup. The wing kinematics
are given by $\phi=\phi_{m}\cos\left(2\pi t\right)$, $\alpha=\frac{\alpha_{m}}{\tanh c_{\alpha}}\tanh\left(c_{\alpha}\sin\left(2\pi t\right)\right)$
and $\theta=0$, where $\phi_{m}=80^{\circ}$, $\alpha_{m}=45^{\circ}$,
$c_{\alpha}=3.3$; the motion is symmetric for the up- and downstroke
and depicted in figure \ref{fig:suzuki_ALL} (a-b). The rectangular
wing and the wing coordinate system are illustrated in figure \ref{fig:suzuki_ALL}
(c). We normalize the distance from pivot to tip to unity, which yields
$a=1.6667$, $b=0.0667$, $B=0.4167$ and a wing thickness of $h=0.04171$.
The Reynolds number is set to $\mathrm{Re}=U_{\mathrm{tip}}B/\nu=100$
with $U_{\mathrm{tip}}=2\pi\phi_{m}$, yielding the kinematic viscosity
$\nu=0.0366$. In the present simulations, we discretize the domain
of size $3\times3\times3$ using $512\times512\times512$ points and
a penalization parameter of $C_{\eta}=1.25\cdot10^{-4}$ ($K=0.365$).
The reference computation in \cite{Suzuki2015} is performed in a
domain of size $4.16\times4.16\times4.16$, using a fine grid near
the wing and a coarse one in the far-field. Based on the resolution
of the fine grid, $\Delta x=B/50$, the corresponding equidistant
resolution would be $500^{3}$. In our simulations, the body reference
point is located at $\underline{x}_{\mathrm{cntr}}^{(g)}=(1.5,\,1.5,\,1.7)$
and coincides with the wing pivot point, thus $\underline{x}_{\mathrm{pivot}}^{(b)}=0$.
The orientation of the body coordinate system is given by $\psi=0^{\circ}$,
$\beta=-45^{\circ}$, $\gamma=45^{\circ}$ and $\eta=-45^{\circ}$,
where the $45^{\circ}$ yaw angle was added to keep the wing as far
away from the vorticity sponges as possible. A total of four strokes
has been computed, starting with a quiescent initial condition, $\underline{u}\left(\underline{x},t=0\right)=0$.
The outer boundary conditions on the domain are homogeneous Dirichlet
conditions in $z$-direction and a vorticity sponge, extending over
$32$ grid points with $C_{\mathrm{sp}}=10^{-1}$, in the remaining
ones. The simulation required 35 GB of memory and 5785 CPU hours on
1024 cores. A total of 27 701 time steps was performed.

The resulting time series of the vertical force is illustrated in
figure \ref{fig:suzuki_ALL} (d). It takes the first two wingbeats
to develop a periodic state, since the motion is impulsively started,
and then the following strokes are almost identical. The present solution
agrees well with the reference solution, and the relative R.M.S difference
is $\left\Vert F-F_{\mathrm{ref}}\right\Vert _{2}/\left\Vert F_{\mathrm{ref}}\right\Vert _{2}\approx4\%$
over the last two periods.

\subsection{Hovering flight of a fruit fly model}

\begin{figure}[t]
\begin{centering}
\includegraphics[width=1\textwidth]{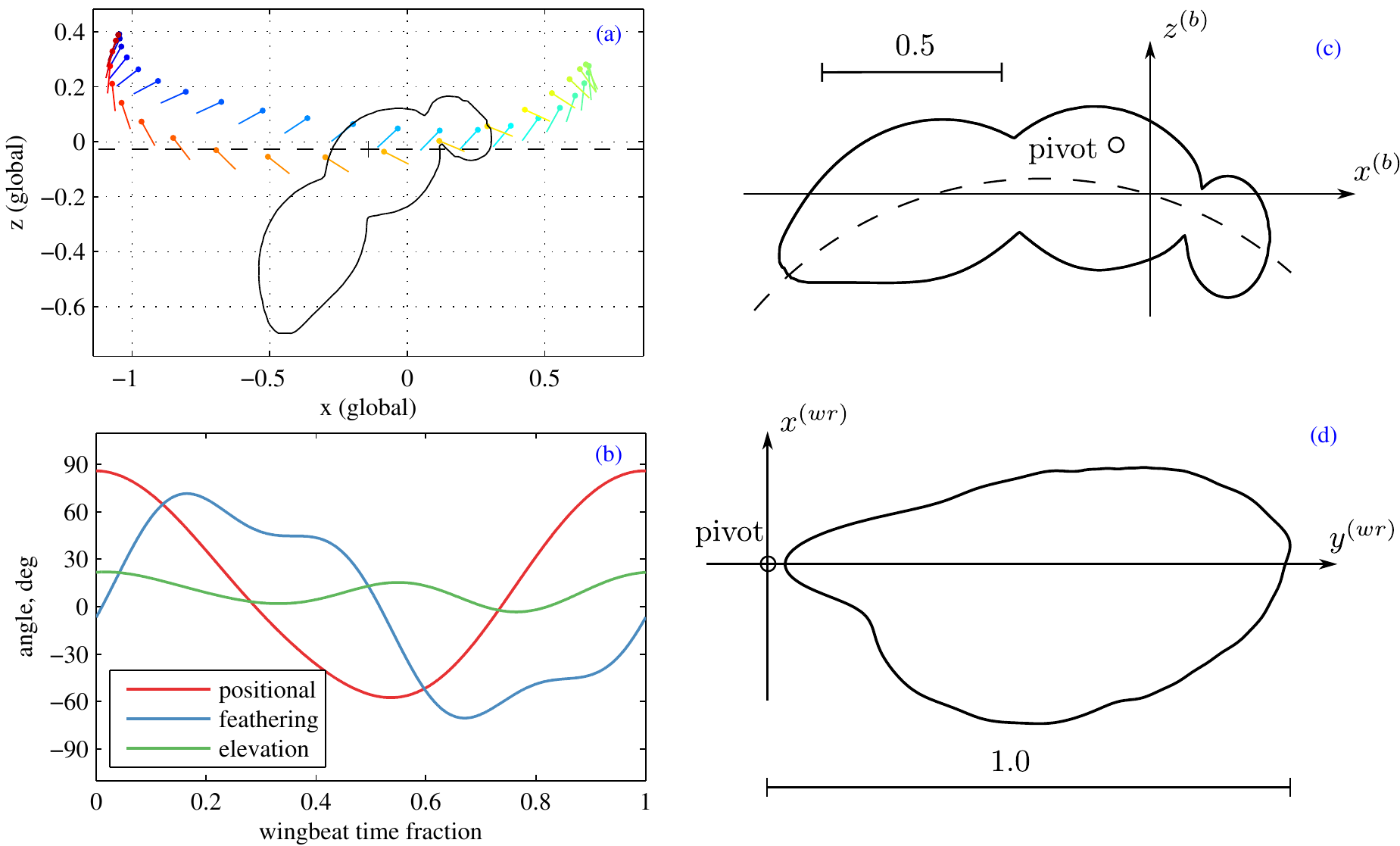}
\par\end{centering}

\caption{Hovering fruit fly, for comparison with \cite{Maeda2013}. (a): wingbeat
kinematics in the side view. (b) time evolution of the wing angles.
(c) Drawing of the insect's body in the plane $y^{(b)}=0$. The body
is rotationally symmetric with a circular center line. (d) Drawing
of the wing shape, together with its pivot point.\label{fig:maeda_geometry_kinematics}}
\end{figure}

The third validation test compares with a hovering fruit fly model,
of which numerical simulations have been presented by Maeda and Liu
\cite{Maeda2013}. Their simulations are based on overset grids, i.e.,
a body-fitted grid for the wings ($65\times65\times11$ each) and
the body ($61\times61\times9$), as well as a background grid ($161\times141\times127$)
have been used to solve the incompressible Navier--Stokes equations,
approximated in the artificial compressibility formulation, using
a finite volume discretization.

The fruit fly considered is defined in figure \ref{fig:maeda_geometry_kinematics}.
The wing length from pivot point to wing tip, $R=2.47\,\mathrm{mm}$,
the fluid density $\varrho_{f}=1.225\,\mathrm{kg/m^{3}}$ and the
wingbeat frequency $f=218\,\mathrm{Hz}$ are used for normalization.
The body, depicted in figure \ref{fig:maeda_geometry_kinematics}
(a,c), has an elliptical cross-section with center points following
an arched centerline of radius $r_{\mathrm{bc}}=0.94644$ centered
at $x_{\mathrm{bc}}^{(b)}=-0.244769$, $z_{\mathrm{bc}}^{(b)}=-0.9301256$.
The wing pivot points are located at $\underline{x}_{\mathrm{pivot,rl}}^{(b)}=\left(-0.12,\,\pm0.1445,\,0.08\right).$
To facilitate reproduction, the supplementary material contains \texttt{STL}
files (StereoLithography) of the fruitfly geometry, as well the parameter
file that can be used to reproduce the results with \texttt{FluSI}.
The insect is tethered at $\underline{x}_{c}^{(g)}=\left(1.6,\,1.6,\,1.9\right)$
in a computational domain of size $3.2\times3.2\times3.2$, discretized
with $640\times640\times640$ Fourier modes and a penalization parameter
of $C_{\eta}=1.15\cdot10^{-4}$ ($K=0.23$). Its body orientation
is given by $\psi=0$, $\beta=-45^{\circ}$, $\gamma=45^{\circ}$
and $\eta=-45^{\circ}$. The fruit fly hovers, thus the body position
and orientation do not change in time. A total of four wing beats
has been computed, and the initial condition is fluid at rest, $\underline{u}\left(\underline{x},t=0\right)=\underline{0}$.
We apply a vorticity sponge in the $x$ and $y$-direction ($32$
grid points thick with $C_{\mathrm{sp}}=10^{-1})$ and impose no-slip
boundary conditions in the $z$-direction, i.e., we impose floor and
ceiling. The simulation required 68 GB of memory and 15 100 CPU hours
on 1024 cores. A total of 48 200 time steps was performed.

The wing shape is illustrated in figure \ref{fig:maeda_geometry_kinematics}
(d). It has a mean chord $c_{m}=A/R=0.33$, which yields with the
kinematic viscosity of air, $\nu=1.5\cdot10^{-5}\,\mathrm{m^{2}/s}$
the Reynolds number $\mathrm{Re}=U_{\mathrm{tip}}c_{m}/\nu=2\Phi Rfc_{m}/\nu=136$
where $\Phi=2.44\,\mathrm{rad}$ is the stroke amplitude of the positional
angle. The time evolution of the wing kinematics is illustrated in
figure \ref{fig:maeda_geometry_kinematics} (b). The up- and downstroke
are not symmetric, and the wing trajectory, figure \ref{fig:maeda_geometry_kinematics}
(a), shows the characteristic $\cup$-shape.

\begin{figure}
\begin{centering}
\includegraphics[width=0.85\textwidth]{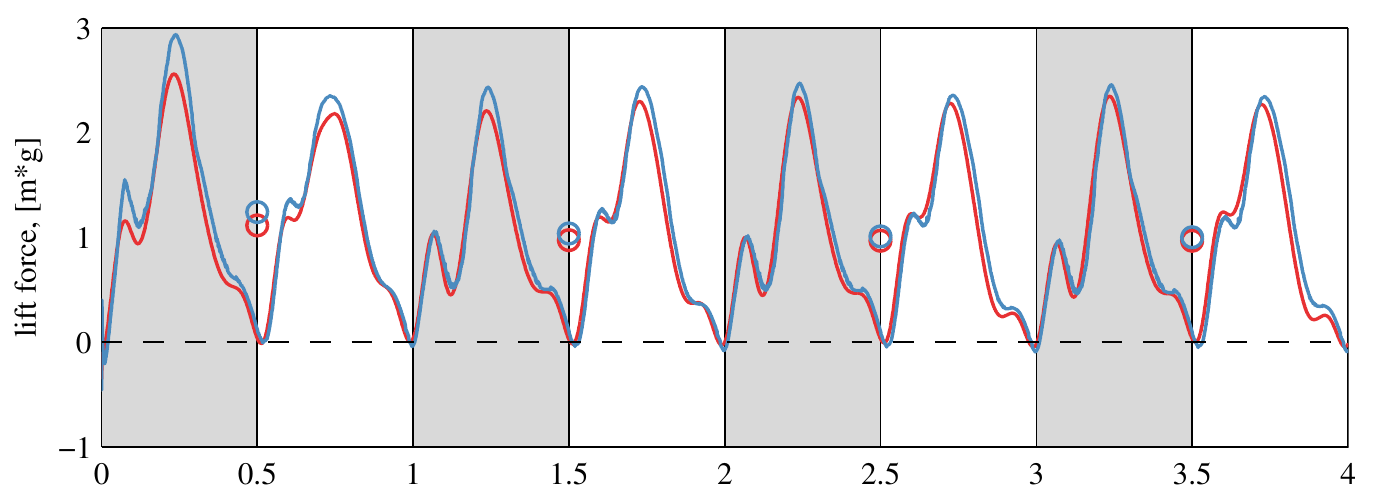}
\par\end{centering}

\begin{centering}
\includegraphics[width=0.85\textwidth]{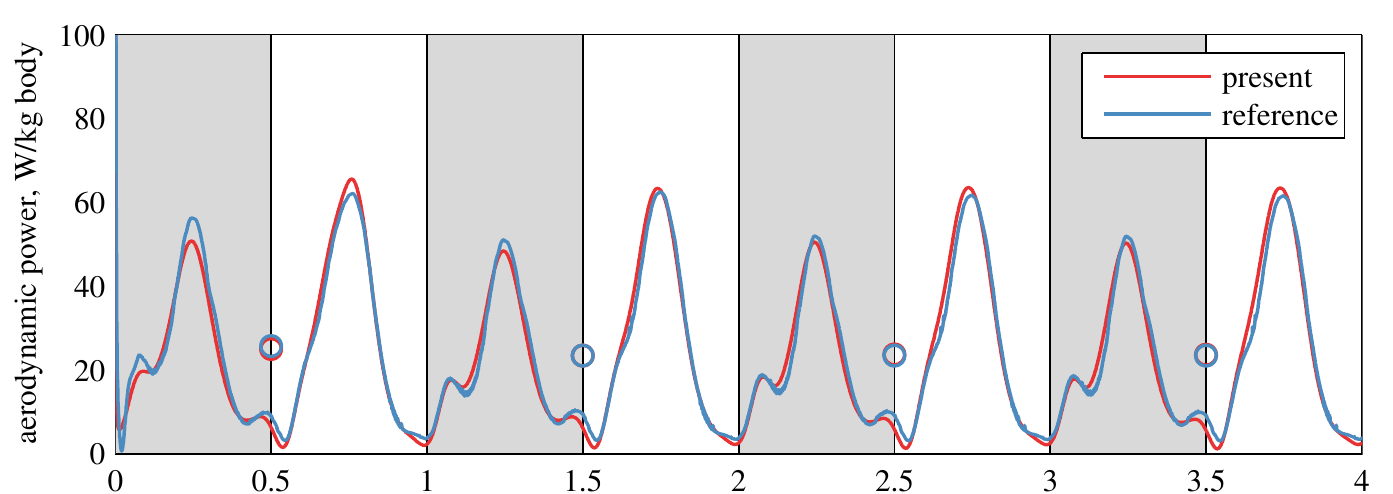}
\par\end{centering}

\caption{Hovering fruit fly, comparison with \cite{Maeda2013}. Top: total
vertical force. Bottom: aerodynamic power. Gray shaded areas denote
upstrokes, stroke averages are indicated by circles. The mean values
during the last stroke differ by 3.26\% and 1.00\% for the vertical
force and the power, respectively. \label{fig:maeda_fruitfly_results}}
\end{figure}

The results for the lift force, normalized with the weight, and the
aerodynamic power in W/kg body mass are presented in figure \ref{fig:maeda_fruitfly_results}.
As the mass of the model insect was not given in \cite{Maeda2013},
we assumed the fourth stroke of the reference data to balance the
weight, yielding $m=1.02\,\mathrm{mg}$. Small circles at mid-stroke
indicate stroke-averaged values. For the stroke-averaged vertical
force, we find respectively $10.03\%$, $6.19\%$, $4.03\%$ and $3.26\%$
relative difference to the reference data for the four strokes computed,
which can be explained by the impulsively started motion at the beginning
of the first stroke. The time evolution, e.g., the occurrence of peaks,
are very similar in both data sets. The agreement is even better for
the aerodynamic power, with relative differences of $0.57\%$, $0.06\%$,
$-0.95\%$ and $-1.00\%$ for the stroke-averaged values. Both power
and lift peak during the translation phase of the up- and downstroke,
and reach a minimum value at the reversals. 

\begin{figure}
\noindent \begin{centering}
\includegraphics[width=0.75\textwidth]{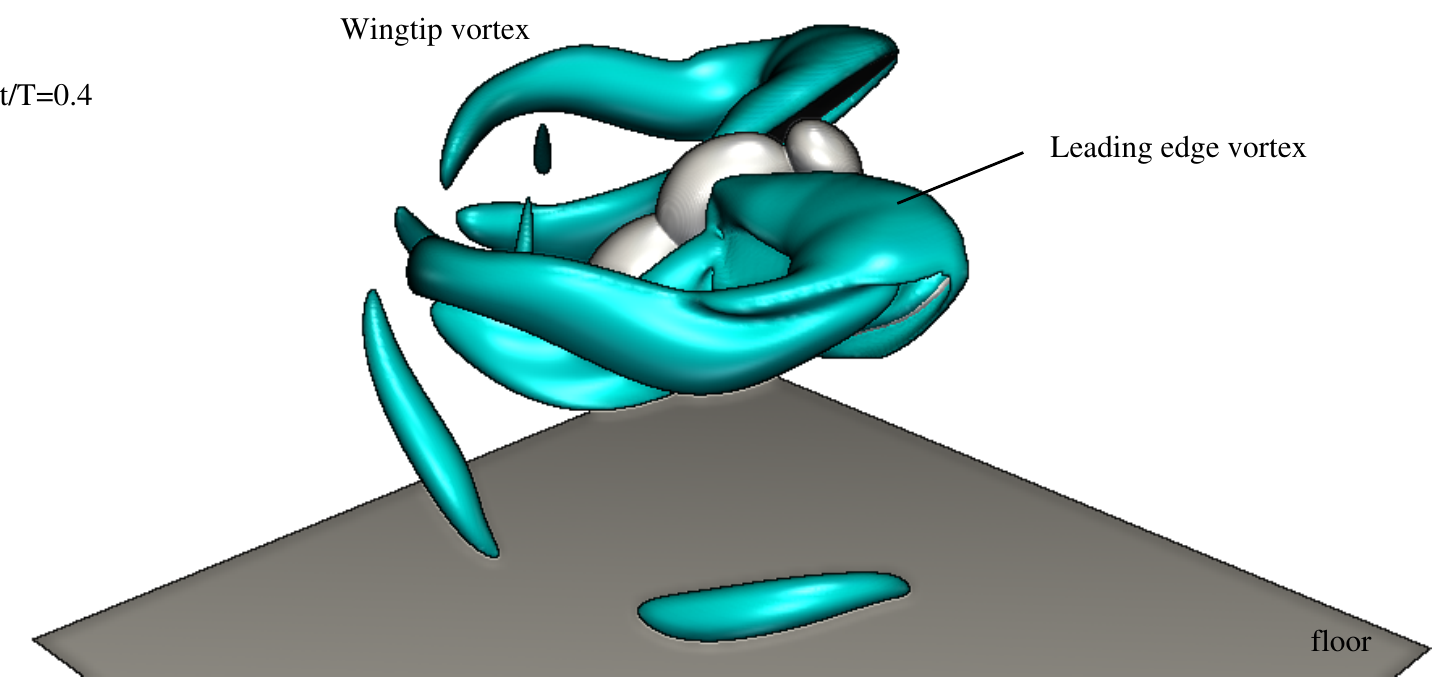}
\par\end{centering}

\caption{\label{fig:maeda_fruitfly_flow_field}Snapshot of the vortical structures
generated by a hovering fruit fly model, visualized by iso-surfaces
of the $Q$-criterion ($Q=100$) shortly before the ventral stroke
reversal $t/T=0.4$. Typical feature of flapping flight, such as the
leading edge and wingtip vortex, are visible. }
\end{figure}

The flow field generated by the fruit fly model is visualized in figure
\ref{fig:maeda_fruitfly_flow_field} by iso-surfaces of the $Q$-criterion,
which can, for incompressible flows, be computed as $Q=\nabla^{2}p/2$,
see \cite[p. 23]{Lesieur2005}. The flow field exhibits the typical
features, such as a leading edge vortex and a wingtip vortex.

\subsection{Butterfly model in free flight}

\begin{figure}
\begin{centering}
\includegraphics[width=0.5\textwidth]{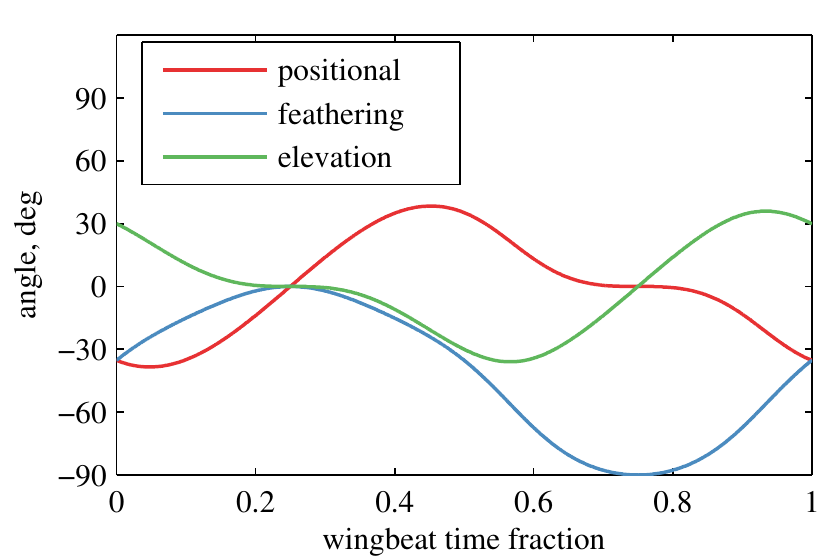}\includegraphics[width=0.5\textwidth]{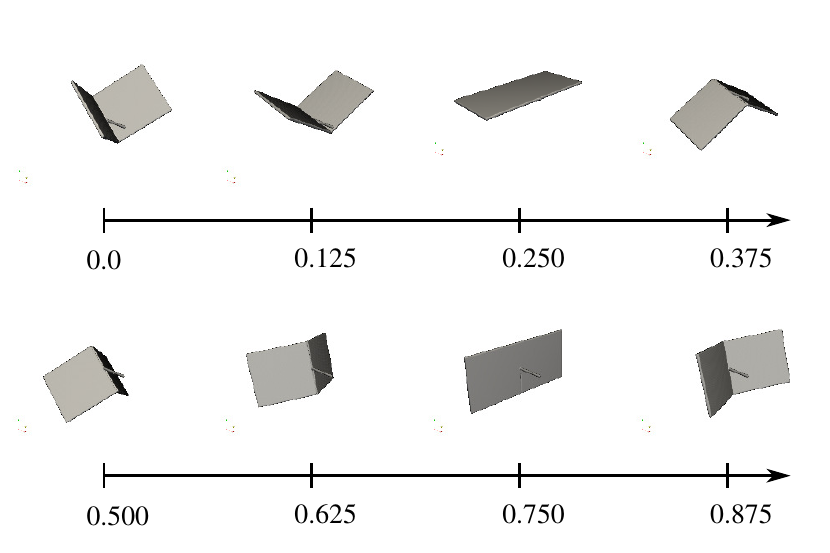}
\par\end{centering}

\begin{centering}
\includegraphics[width=0.5\textwidth]{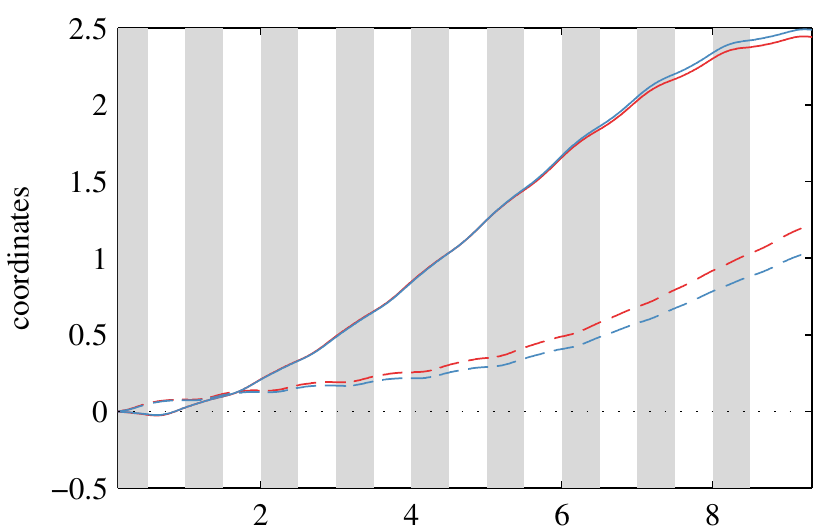}\includegraphics[width=0.5\textwidth]{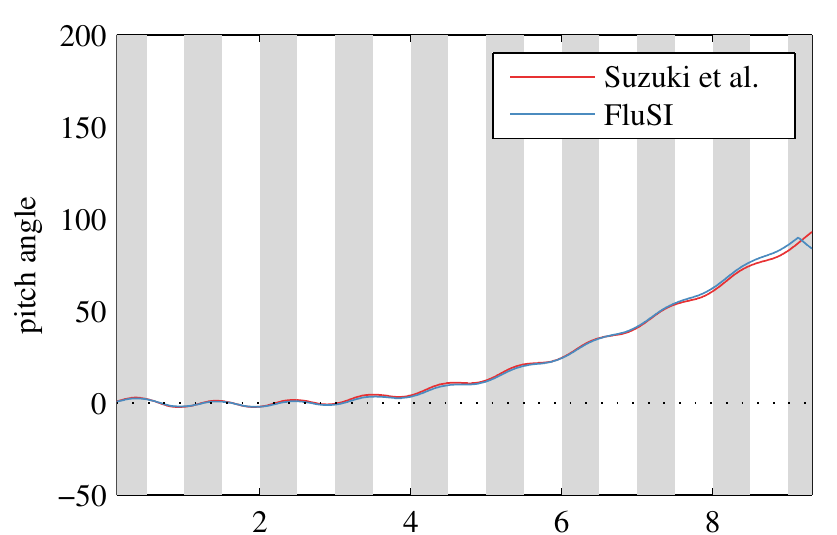}
\par\end{centering}

\begin{centering}
\includegraphics[width=0.5\textwidth]{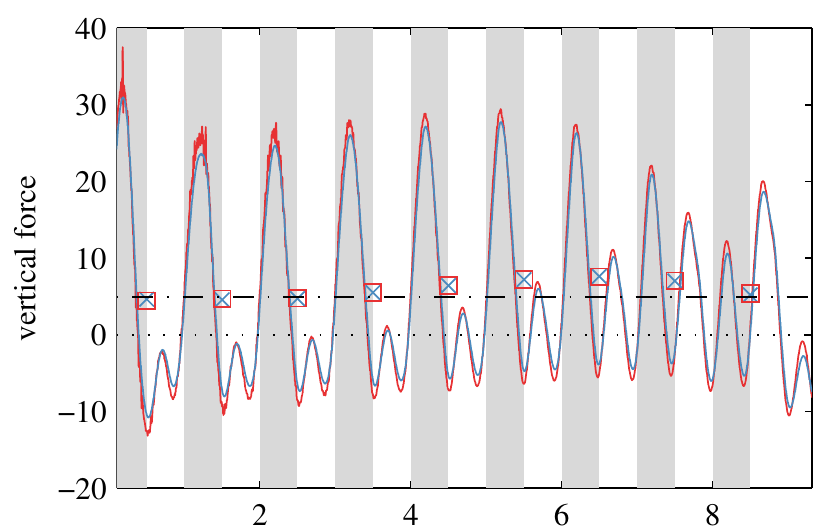}\includegraphics[width=0.5\textwidth]{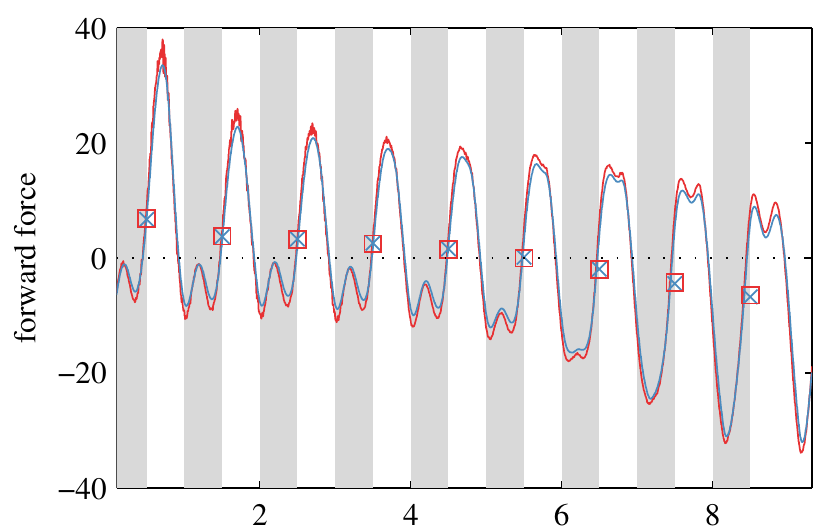}
\par\end{centering}

\caption{Freefly flying butterfly\label{fig:Freefly-flying-butterfly} model
for comparison with \cite{Suzuki2015}. Top: kinematics and illustration
of the wingbeat. Mid left: coordinates of the center of mass for the
horizontal (solid lines) and vertical (dashed line) component. Mid
right: body pitch angle $\beta$. Bottom: vertical (left) and forward
force (right) component. Symbols indicate stroke averages. Dash-dotted
line marks weight. Overall agreement is good.}

\end{figure}

To complete the validation section, we consider the freely flying
butterfly model presented in \cite[section 5.3]{Suzuki2015}. The
body consists of a thin rod with quadratic wings of length $R=1$,
and the wing kinematics is inspired by a butterfly. Figure \ref{fig:Freefly-flying-butterfly}
(top row) illustrates the prescribed wingbeat kinematics, converted
to the convention used in this work. The Reynolds number is $Re=300=U_{\mathrm{tip}}R/\nu$,
where $U_{\mathrm{tip}}=\pi$ is the mean wingtip velocity. The mass
of the butterfly is $M=38$, and the moments of inertia of the body
are given by $J_{y}^{(b)}=J_{z}^{(b)}=3.1667$. The roll motion is
inhibited (five degrees of freedom) and the mass of the wings is neglected.
Gravitational acceleration is given by $g=0.1304$. Present computations
are performed on a domain of size $6\times6\times6$ using a resolution
of $1024\times1024\times1024$ and the penalization constant is $C_{\eta}=1.4\cdot10^{-4}$
($K=0.2$). We apply a vorticity sponge in all directions, centered
around the moving center of the insect $\underline{x}_{\mathrm{cntr}}^{(g)}$.
We computed 9 strokes (65066 time steps), the computation allocated
$200$ GB of memory and consumed $79915$ CPU hours on $4096$ cores.
The parameter files used in this compuation are in the supplementary
material.

Figure \ref{fig:Freefly-flying-butterfly} (mid row) compares the
coordinates of the center of mass and the body pitch angle with the
reference solution, showing good agreement with the reference computation.
A slight deviation in the vertical component can be observed, which
can be explained by the fact that the slight differences in the forces
(Fig. \ref{fig:Freefly-flying-butterfly} bottom row) are integrated
twice with respect to time. The general agreement is good, and we
conclude that our code is validated.

\section{\label{sec:Application-to-a}Application to a bumblebee model in
forward flight}

In order to demonstrate the applicability of the software environment
to more complex insects at higher Reynolds number, we consider in
this chapter a different insect model with a Reynolds number of $\mathrm{Re}=2\Phi Rfc_{m}/\nu=2042$.
The model is derived from a bumblebee and its detailed morphology
and kinematics can be found in the supplementary material (\texttt{STL}
file of the geometry as well as the parameter files that can be used
to reproduce the results with \texttt{FluSI}). The model has also
been used in \cite[suppl. mat.]{Engels2015}. The body contains details
such as the legs and antennae, since they can be easily included in
the present framework. The bumblebee is considered in forward flight
at $u_{\infty}=1.246$. The domain size is set to $6\times4\times4$
and discretized with $1152\times768\times768$ grid points. The penalization
parameter is set to $C_{\eta}=2.5\cdot10^{-4}$ ($K=0.074$). Four
strokes have been simulated, requiring $26467$ CPU hours on 1024
cores and 153.6 GB of memory.

\begin{figure}
\begin{centering}
\includegraphics[width=1\textwidth]{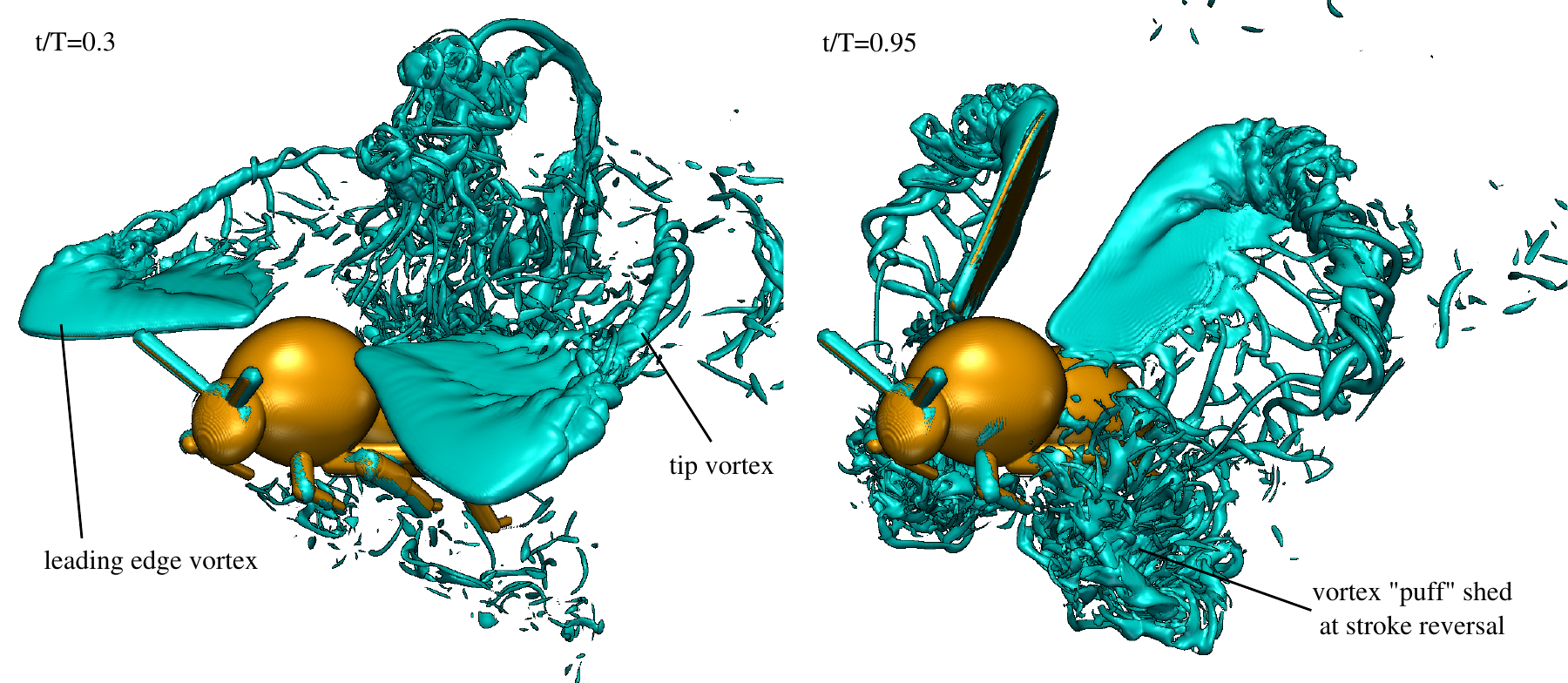}
\par\end{centering}

\caption{\label{fig:BB-Flow-field}Flow field generated by the model bumblebee
in forward flight, visualized by the $\left\Vert \underline{\omega}\right\Vert =100$
isosurface of vorticity magnitude. }
\end{figure}

\begin{figure}
\begin{centering}
\includegraphics[width=1\textwidth]{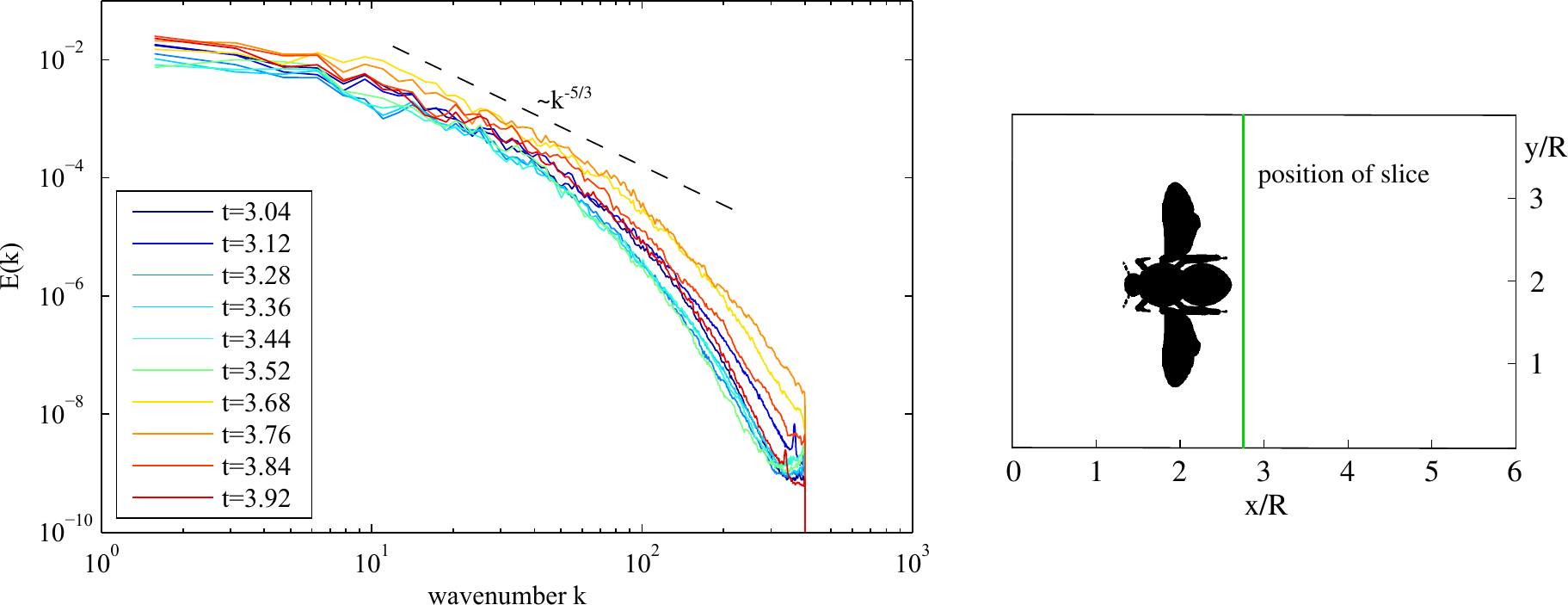}
\par\end{centering}

\caption{\label{fig:BB-Spectra}Right: Radial energy spectra in a slice perpendicular
to the flow direction ($x=2.67$). The spectra are full and exhibit
an inertial- and dissipative range. The resolution is $1152\times768\times768$,
thus the highest resolved wavenumber is $k_{\mathrm{max}}=256\cdot\left(2\pi/4\right)$,
due to the de-aliasing using the $2/3$ rule. Right: position of the
slice and insect in the computational domain of size $6\times4\times4$.}

\end{figure}

Figure \ref{fig:BB-Flow-field} visualizes the flow field generated
by the model by the $\left\Vert \underline{\omega}\right\Vert =100$
isosurface of vorticity magnitude, for the times $t/T=0.3$ and $0.95$,
where $T$ is the period time. At $t/T=0.3$, the leading edge vortex
and the wingtip vortex are clearly visible, yet the flow field presents
much smaller scales than in the case of a fruitfly (Fig. \ref{fig:maeda_fruitfly_flow_field}).
At $t/T=0.95$, which is shortly before the beginning of a new cycle,
the vortex ``puff'' shed at the stroke reversal ($t/T\approx0.5$)
are visible. The flow field is both spatially and temporally intermittent.
Figure \ref{fig:BB-Flow-field} gives the impression of a turbulent
flow field, which can be quantified by the energy spectra of the
velocity in a 2D slice perpendicular to the flow direction. Figure
\ref{fig:BB-Spectra} shows the chosen position of the slice at $x=2.67$,
and the radial energy spectra for 10 times throughout the cycle.
 At any time $t/T$, the energy spectrum $E(k)=1/2\sum_{k-1/2\le|\underline{k}|<k+1/2}\left|\widehat{u}_{x}\left(\underline{k}\right)\right|{}^{2}+\left|\widehat{u}_{z}\left(\underline{k}\right)\right|^{2}+\left|\widehat{u}_{z}\left(\underline{k}\right)\right|^{2}$,
where $\underline{k}=\left(k_{x},k_{y}\right)$, is full and exhibits
an intertial range with a $k^{-5/3}$ slope, which is a typical feature
of turbulence.

\begin{figure}
\begin{centering}
\includegraphics[width=0.5\textwidth]{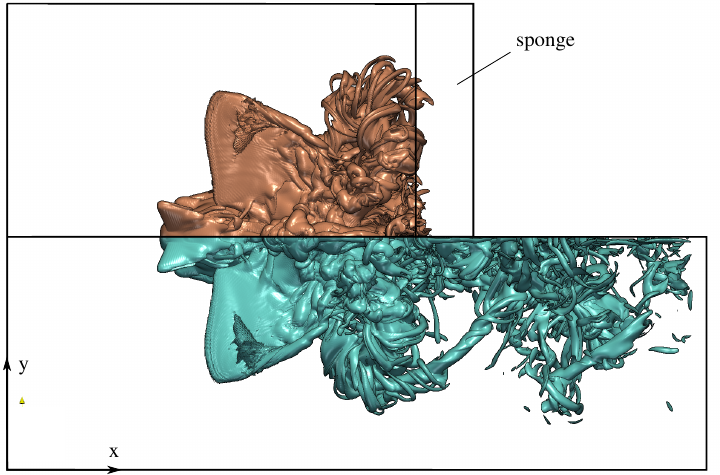}
\par\end{centering}

\caption{\label{fig:Influence-sponge}Influence of the vorticity sponge in
a bumblebee simulation. Shown is the $\left\Vert \underline{\omega}\right\Vert =20$
isosurface of vorticity magnitude. A simulation with a longer ($\ell_{x}=6$,
blue) and a shorter domain ($\ell_{x}=4$, orange) have been performed.
In the shorter simulation, the sponge occupies the region $3.5\leq x\leq4$.
The sponge constant is $C_{\mathrm{sp}}=10^{-1}.$}
\end{figure}

As described in section \ref{sub:Wake-removal-techniques}, a vorticity
sponge technique has been used to overcome the periodicity. Let us
briefly discuss the influence of the sponge and the choice of parameters.
Figure \ref{fig:Influence-sponge} visualizes the flow field from
two simulations, one with a shorter domain ($4\times4\times4$) and
the previously discussed one. In the shorter domain, a sponge is active
in the region $3.5\leq x\leq4$ with a constant of $C_{\mathrm{sp}}=10^{-1}$.

\section{Conclusions and Perspectives\label{sec:Conclusions-and-Perspectives}}

This article presents the open source software \texttt{FluSI} (\href{https://github.com/pseudospectators/FLUSI}{https://github.com/pseudospectators/FLUSI})
for the numerical simulation of the aerodynamics of flapping insect
flight running on massive parallel computer architectures. Different
benchmarks demonstrated the efficiency of the code and showed its
validity in comparison with results from the literature. The numerical
method is based on a Fourier pseudo-spectral discretization of the
three-dimensional incompressible Navier--Stokes equations. Thus no
artificial numerical diffusion or dispersion are introduced in the
discretization. The no-slip boundary conditions for the complexly
shaped and time varying geometry of the flapping wings and the insect
body are imposed with the volume penalization method. The penalization
parameter is chosen such that the modeling error, due to penalization,
and the discretization error are balanced. The computational cost
of the flow solver is essentially due to the Fourier transforms. Benefiting
from the efficient implementation of three-dimensional FFT, excellent
scaling on large scale computing clusters has thus been obtained.
A limitation of the current approach is that equidistant Cartesian
grids are required and a compromize between the domain size and the
number of grid points has to be imposed. The modular structure of
the\texttt{ FluSI} code permits to design different and complex geometries
of the insect shape and its wings easily and also to change their
kinematics. For the free flight option, equations in a quaternion-based
formulation are solved. Different flow configurations, like channel
flows with laminar or turbulent inflow or including bluff bodies of
almost arbitrary shape, are possible using penalization together with
a sponge technique. 

\bibliographystyle{siam}
\phantomsection\addcontentsline{toc}{section}{\refname}\bibliography{/home/engels/Documents/Research/Articles/bibliography}

\begin{thebibliography}{10}

\bibitem{Angot1999}
{\sc P.~Angot, C.~Bruneau, and P.~Fabrie}, {\em A penalization method to take
  into account obstacles in incompressible viscous flows}, Numer. Math., 81
  (1999), pp.~497--520.

\bibitem{Arquis1984}
{\sc E.~Arquis and J.-P. Caltagirone}, {\em Sur les conditions hydrodynamiques
  au voisinage d'une interface milieu fluide milieu poreux: application \`a la
  convection naturelle}, C. R. Acad. Sci. Paris, S\'er. II, 299 (1984).

\bibitem{Berman2007}
{\sc G.~J. Berman and Z.~J. Wang}, {\em Energy-minimizing kinematics in
  hovering insect flight}, J. Fluid Mech., 582 (2007), pp.~153--168.

\bibitem{Bimbard2013}
{\sc G.~Bimbard, D.~Kolomenskiy, O.~Bouteleux, J.~Casas, and R.~Godoy-Diana},
  {\em Force balance in the take-off of a pierid butterfly: relative importance
  and timing of leg impulsion and aerodynamic forces}, J. Exp. Biol., 216
  (2013), pp.~3551--3563.

\bibitem{Zang1986}
{\sc C.~Canuto, M.~Y. Hussaini, A.~Quarteroni, and T.~Zang}, {\em Spectral
  Methods in Fluid Dynamics}, Springer Verlag, 1986.

\bibitem{Carbou2003}
{\sc G.~Carbou and P.~Fabrie}, {\em Boundary layer for a penalization method
  for viscous incompressible flow}, Adv. Diff. Equ., 8 (2003), pp.~1453--2480.

\bibitem{Cooley1965}
{\sc J.~W. Cooley and J.~W. Tukey}, {\em An algorithm for the machine
  calculation of complex {F}ourier series}, Math. Comput., 19 (1965),
  pp.~297--301.

\bibitem{Engels2012a}
{\sc T.~Engels, D.~Kolomenskiy, K.~Schneider, and J.~Sesterhenn}, {\em
  Two-dimensional simulation of the fluttering instability using a
  pseudospectral method with volume penalization}, Computers \& Structures, 122
  (2012), pp.~101--112.

\bibitem{Engels2014}
{\sc T.~Engels, D.~Kolomenskiy, K.~Schneider, and J.~Sesterhenn}, {\em
  Numerical simulation of fluid-structure interaction with the volume
  penalization method}, J. Comput. Phys., 281 (2015), pp.~96--115.

\bibitem{Engels2015}
{\sc T.~Engels, D.~Kolomenskiy, K.~Schneider, J.~Sesterhenn, and F.-O.
  Lehmann}, {\em Bumblebee flight in heavy turbulence}, Phys. Rev. Lett.,
  accepted,  (2015).

\bibitem{Frigo2005}
{\sc M.~Frigo and S.~G. Johnson}, {\em The design and implementation of
  {FFTW3}}, Proc. IEEE, 94 (2005), pp.~216--231.

\bibitem{Hejlesen2015}
{\sc M.~Hejlesen, P.~Koumoutsakos, A.~Leonard, and J.~Walther}, {\em Iterative
  {B}rinkman penalization for remeshed vortex methods.}, J. Comput. Phys., 280
  (2015), pp.~547--562.

\bibitem{Introini2014}
{\sc C.~Introini, M.~Belliard, and C.~Fournier}, {\em A second order penalized
  direct forcing for hybrid {C}artesian/immersed boundary flow simulations},
  Computers \& Fluids, 90 (2014), pp.~21--41.

\bibitem{Ji2012}
{\sc C.~Ji, A.~Munjiza, and J.~Williams}, {\em A novel iterative direct-forcing
  immersed boundary method and its finite volume applications}, J. Comput.
  Phys., 231 (2012), pp.~1797--1821.

\bibitem{Kadoch_etal_2012}
{\sc B.~Kadoch, D.~Kolomenskiy, P.~Angot, and K.~Schneider}, {\em A volume
  penalization method for incompressible flows and scalar advection-diffusion
  with moving obstacles}, J. Comput. Phys., 231 (2012), pp.~4365--4383.

\bibitem{Kim1985}
{\sc J.~Kim and P.~Moin}, {\em Application of a fractional-step method to
  incompressible navier-stokes equations}, J. Comput. Phys., 59 (1985),
  pp.~308--323.

\bibitem{Kolomenskiy2009}
{\sc D.~Kolomenskiy and K.~Schneider}, {\em A {F}ourier spectral method for the
  {N}avier-{S}tokes equations with volume penalization for moving solid
  obstacles}, J. Comput. Phys., 228 (2009), pp.~5687--5709.

\bibitem{Lesieur2005}
{\sc M.~Lesieur, O.~Métais, and P.~Comte}, {\em Large-Eddy Simulations of
  Turbulence}, Cambridge University Press, 2005.

\bibitem{Liu2009}
{\sc H.~Liu}, {\em Integrated modeling of insect flight: From morphology,
  kinematics to aerodynamics}, J. Comput. Phys., 228 (2009), pp.~439--459.

\bibitem{Maeda2010}
{\sc M.~Maeda, N.~Gao, N.~Nishihashi, and H.~Liu}, {\em A free-flight
  simulation of insect flapping flight}, J. Aero Aqua Bio-mech., 1 (2010),
  pp.~71--79.

\bibitem{Maeda2013}
{\sc M.~Maeda and H.~Liu}, {\em Ground effect in fruit fly hovering: A
  three-dimensional computational study.}, J. Biomech. Sc. Engin., 8 (2013),
  pp.~344--355.

\bibitem{Mittal2005}
{\sc R.~Mittal and G.~Iaccarino}, {\em Immersed boundary methods}, Annu. Rev.
  Fluid Mech., 37 (2005), pp.~239--261.

\bibitem{Mordant2000}
{\sc N.~Mordant and J.-F. Pinton}, {\em Velocity measurement of a settling
  sphere}, Eur. Phys. J. B, 18 (2000), pp.~343--352.

\bibitem{Romain2012}
{\sc R.~{Nguyen van yen}, D.~Kolomenskiy, and K.~Schneider}, {\em Approximation
  of the {L}aplace and {S}tokes operators with {D}irichlet boundary conditions
  through volume penalization: A spectral viewpoint}, Numer. Math., 128 (2014),
  pp.~301--338.

\bibitem{Osher2003}
{\sc S.~Osher and R.~Fedkiw}, {\em Level Set Methods and Dynamic implicit
  Surfaces}, Springer, 2003.

\bibitem{Pekurovsky2012}
{\sc D.~Pekurovsky}, {\em {P3DFFT}: a framework for parallel computations of
  {F}ourier transforms in three dimensions}, SIAM J. Sci. Comput., 34 (2012),
  pp.~C192--C209.

\bibitem{Peskin2002}
{\sc C.~S. Peskin}, {\em The immersed boundary method}, Acta Numerica, 11
  (2002), pp.~479--517.

\bibitem{Peyret2002}
{\sc R.~Peyret}, {\em Spectral Methods for Incompressible Viscous Flow},
  Springer Berlin / Heidelberg, 2002.

\bibitem{Ramamurti2009}
{\sc R.~Ramamurti and W.~Sandberg}, {\em A computational investigation of the
  three-dimensional unsteady aerodynamics of drosophila hovering and
  maneuvering}, J. Exp. Biol., 210 (2009), pp.~881--896.

\bibitem{Schlatter2005}
{\sc P.~Schlatter, N.~Adams, and L.~Kleiser}, {\em A windowing method for
  periodic inflow/outflow boundary treatment of non-periodic flows}, J. Comput.
  Phys., 206 (2005), pp.~505--535.

\bibitem{Schneider2005}
{\sc K.~Schneider}, {\em Numerical simulation of the transient flow behaviour
  in chemical reactors using a penalisation method}, Computers \& Fluids, 34
  (2005), pp.~1223--1238.

\bibitem{Schneider2015}
{\sc K.~Schneider}, {\em Immersed boundary methods for numerical simulation of
  confined fluid and plasma turbulence in complex geometries: a review.}, J.
  Plasma Phys., 81 (2015), p.~435810601.

\bibitem{Schneider2005a}
{\sc K.~Schneider and M.~Farge}, {\em Numerical simulation of the transient
  flow behaviour in tube bundles using a volume penalization method}, J. Fluids
  Struct., 20 (2005), pp.~555--566.

\bibitem{Suzuki2015}
{\sc K.~Suzuki, K.~Minami, and T.~Inamuro}, {\em Lift and thrust generation by
  a butterfly-like flapping wing-body model: immersed boundary-lattice
  {B}oltzmann simulations}, J. Fluid Mech., 767 (2015), pp.~659--695.

\bibitem{Thompson1985}
{\sc J.~F. Thompson, Z.~Warsi, and C.~W. Mastin}, {\em Numerical grid
  generation: Foundations and Applications.}, North-Holland Amsterdam, 1985.

\bibitem{Uhlmann2005}
{\sc M.~Uhlmann}, {\em An immersed boundary method with direct forcing for the
  simulation of particulate flows}, J. Comput. Phys., 209 (2005), pp.~448--476.

\end{thebibliography}

\end{document}